\documentclass[lettersize,journal]{IEEEtran}
\usepackage{xcolor}
\usepackage{acronym}
\usepackage{graphicx}
\usepackage{amsmath}
\usepackage{tabularx,booktabs}
\usepackage{multicol}
\usepackage{makecell}
\usepackage{amsfonts}
\usepackage{bm}
\usepackage{tablefootnote}
\usepackage{subcaption}
\usepackage{url}
\usepackage{caption}
\usepackage{amssymb}
\usepackage{amsthm}
\usepackage{chngcntr}
\newcolumntype{Y}{>{\centering\arraybackslash}X}
\newcolumntype{?}{!{\vrule width 1.5pt}}
\usepackage[numbers,sort&compress]{natbib}
\acrodef{PMF}[PMF]{probability mass function}
\acrodef{MAC}[MAC]{medium access control}
\acrodef{IRSA}[IRSA]{irregular repetition slotted ALOHA}
\acrodef{i.i.d.}[i.i.d.]{independent and identically distributed}
\acrodef{DL}[DL]{deep learning}
\acrodef{DNN}[DNN]{deep neural network}
\acrodef{BN}[BN]{Bayesian network}
\acrodef{MC}[MC]{Markov chain}
\acrodef{MAB}[MAB]{multi-armed bandit}
\acrodef{IIoT}[IIoT]{industrial Internet-of-things}
\acrodef{URLLC}[URLLC]{ultra-reliable low-latency communication}
\acrodef{MARL}[MARL]{multi-agent reinforcement learning}
\acrodef{RL}[RL]{reinforcement learning}
\acrodef{mMTC}[mMTC]{massive machine-type communication}
\acrodef{ACK}[ACK]{acknowledgment}
\acrodef{CTDE}[CTDE]{centralized-training / decentralized-execution}
\acrodef{GNN}[GNN]{graph neural network}
\acrodef{ST-GNN}[ST-GNN]{spatio-temporal graph neural network}
\acrodef{THz}[THz]{terahertz}
\acrodef{LOS}[LOS]{line-of-sight}
\acrodef{NLOS}[NLOS]{non-line-of-sight}
\acrodef{RRM}[RRM]{radio resource management}
\acrodef{AirGNN}[AirGNN]{over-the-air graph neural network}
\acrodef{AirComp}[AirComp]{over-the-air computation}
\acrodef{mmWave}[mmWave]{millimeter-wave}
\acrodef{RIS}[RIS]{reflective intelligent surface}
\acrodef{CSI}[CSI]{channel state information}
\acrodef{AP}[AP]{access point}
\acrodef{TDD}[TDD]{time division duplexing}
\acrodef{RSS}[RSS]{radio signal strength}
\acrodef{BCE}[BCE]{binary cross-entropy}
\acrodef{MPNN}[MPNN]{message passing neural network}
\acrodef{MIMO}[MIMO]{multiple-input multiple-output}
\acrodef{RIS}[RIS]{reconfigurable intelligent surfaces}
\acrodef{FL}[FL]{federated learning}
\acrodef{DRL}[DRL]{deep reinforcement learning}
\acrodef{VoI}[VoI]{value of information}
\acrodef{AoI}[AoI]{age of information}
\acrodef{GO-ST-AirGNN}[GO-ST-AirGNN]{goal-oriented spatio-temporal over-the-air graph neural network}
\acrodef{ISAC}[ISAC]{integrated sensing and communication}
\acrodef{RF}[RF]{radio frequency}
\acrodef{RNN}[RNN]{recurrent neural network}
\acrodef{SNR}[SNR]{signal-to-noise ratio}
\acrodef{CTDE}[CTDE]{centralized-training decentralized-execution}
\acrodef{AMR}[AMR]{autonomous mobile robot}
\acrodef{LSTM}[LSTM]{long short term memory}
\acrodef{MLP}[MLP]{multi-layer perceptron}
\acrodef{ST-GCN}[ST-GCN]{spatio-temporal graph convolutional network}
\acrodef{ST-GAT}[ST-GAT]{spatio-temporal graph attention network}
\acrodef{QoS}[QoS]{quality-of-service}
\acrodef{TDMA}[TDMA]{time division multiple access}
\acrodef{OFDMA}[OFDMA]{orthogonal frequency division multiple access}
\acrodef{KPI}[KPI]{key performance indicator}

\theoremstyle{definition}

\usepackage{algorithm}
\usepackage{algpseudocode}
\algnewcommand{\LeftComment}[1]{\Statex \(\triangleright\) \textit{#1}}

\usepackage{soul,color}

\usepackage{ragged2e}
\usepackage{blindtext}

\newcommand{\G}{\mathcal{G}}

\newcommand{\Edges}{\mathcal{E}}

\begin{document}

\title{Goal-Oriented Learning at the Edge: Graph Neural Networks Over-the-Air for Blockage Prediction}

\author{Lorenzo~Mario~Amorosa,~\IEEEmembership{Member,~IEEE},~Zhan~Gao,~\IEEEmembership{Graduate~Student~Member,~IEEE},~Tony~Chahoud,\\~\IEEEmembership{Member,~IEEE},~Yiqun~Wu,~Lukas~Eller,~Marco~Skocaj,~Roberto~Verdone,~\IEEEmembership{Senior Member,~IEEE}
%<-this % stops a space

\thanks{
\indent L.M. Amorosa and R. Verdone are with the Department of Electrical, Electronic and Information Engineering (DEI), ``Guglielmo Marconi", University of Bologna \& WiLab - National Wireless Communication Laboratory (CNIT), Bologna, Italy. E-mail: \{lorenzomario.amorosa, roberto.verdone\}@unibo.it\\
\indent Z. Gao is with the Department of Computer Science and Technology, University of Cambridge, Cambridge, U.K. E-mail: zg292@cam.ac.uk\\
\indent T. Chahoud is with WiLab - National Wireless Communication Laboratory (CNIT), Bologna, Italy. E-mail: tony.chahoud@wilab.cnit.it\\
\indent Y. Wu is with Huawei Technologies Co., Ltd., RAN Research Department, Shanghai, China. E-mail: wuyiqun@huawei.com\\
\indent L. Eller and M. Skocaj are with Huawei Heisenberg Research Center (Munich), Germany. E-mail: \{lukas.eller,marco.skocaj\}@huawei.com\\
\indent Corresponding author: Zhan Gao.
}
}

% header
\markboth{}{}

% make the title area
\maketitle

\begin{abstract}
Sixth-generation (6G) wireless networks evolve from connecting devices to connecting intelligence. The focus turns to \emph{Goal-Oriented Communications}, where the effectiveness of communication is assessed through task-level objectives over traditional throughput-centric metrics. As communication intertwines with learning at the edge, distributed inference over wireless networks faces a critical trade-off between task accuracy and efficient radio resource use. Traditional communication schemes (e.g., OFDMA) are not designed for this trade-off, often facing challenges related to scalability and latency. Therefore, we propose a novel goal-oriented framework that integrates over-the-air computation with spatio-temporal graph learning. Leveraging the wireless channel as an analog aggregation layer, the proposed framework enables low-latency message passing while efficiently aggregating semantically relevant features from distributed nodes. Theoretical analysis confirms that our analog architecture converges to the expressive power of digital message passing, while offering decisive scalability advantages. We assess the framework in proactive line-of-sight blockage prediction for millimeter-wave networks. Through high-fidelity ray-tracing simulations, the framework exhibits strong inductive generalization to unseen networks and adapts to domain shifts via lightweight transfer learning, matching or even outperforming digital baselines with significantly reduced communication overhead.
\end{abstract}


\IEEEpeerreviewmaketitle

\section{Introduction}
\label{sec:introduction}

\IEEEPARstart{T}{he} evolution of wireless networks toward sixth generation (6G) is driving a fundamental paradigm shift from purely connecting devices to connecting intelligence \cite{letaief2019roadmap, 9567793}. The rapid proliferation of artificial intelligence (AI)-driven services and applications, combined with the envisioned role of 6G as an enabler of distributed intelligence at the network edge  \cite{saad2019vision} necessitates a change in network design: rather than simply optimizing for bit-rates or packet error rates, the focus turns to the effect of communication in the expected improvement of a task-specific objective \cite{Gunduz2021Effective}.
This necessitates \emph{Goal-Oriented Communications} \cite{pezone2022goal}, where the design of communications schemes becomes an integral component of learning algorithms at the edge, and joint optimization of communication and learning aims to fulfill specific downstream tasks, such as control, anomaly detection, or sensing, rather than merely targeting generic \ac{QoS} metrics \cite{wen2023task}. %, 10597087}.

However, effective deployments of distributed inference over wireless networks, especially in \ac{IIoT} settings, face a critical trade-off between task accuracy and efficient radio resource use. On one hand, robust decision-making typically requires cooperation among nodes to fuse information from multiple spatially distributed nodes \cite{ren2023survey, 9170818}. On the other hand, the resulting predictions must be produced within strict deadlines to enable timely control actions and avoid system failures \cite{10843389}. 
Standard communication systems are designed to maximize the reliable delivery of individual data streams, employing orthogonal resource allocation (e.g., \ac{TDMA}, \ac{OFDMA}) to isolate users and eliminate interference.
While robust, this creates a scalability barrier: in dense industrial networks, the finite availability of radio resources limits the number of devices that can coordinate for a cooperative task through simultaneous transmissions. As a result, the latency incurred to collect reliable data from all relevant sensors often exceeds application requirements, making the inference obsolete \cite{10226176}.

\begin{figure*}[t]
    \centering
    \begin{subfigure}[b]{0.32\textwidth}
        \includegraphics[width=\linewidth]{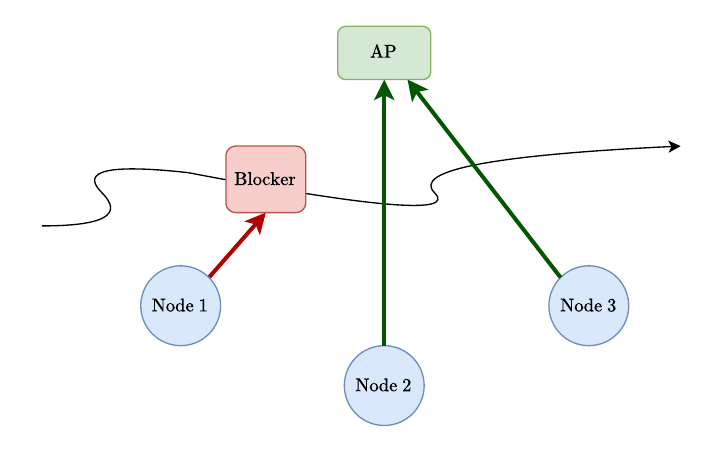}
        \caption{$t=0$: Node 1 detects a blockage.}
    \end{subfigure}
    \hfill
    \begin{subfigure}[b]{0.32\textwidth}
        \includegraphics[width=\linewidth]{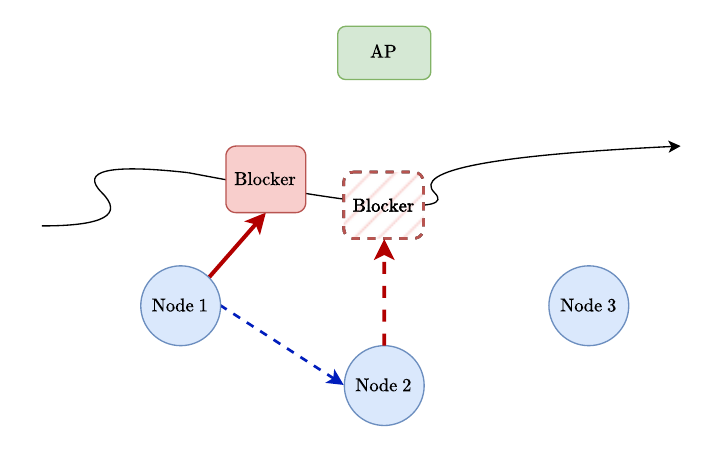}
        \caption{$\tau=t+1$: Node 1 aids Node 2.}
    \end{subfigure}
    \hfill
    \begin{subfigure}[b]{0.32\textwidth}
        \includegraphics[width=\linewidth]{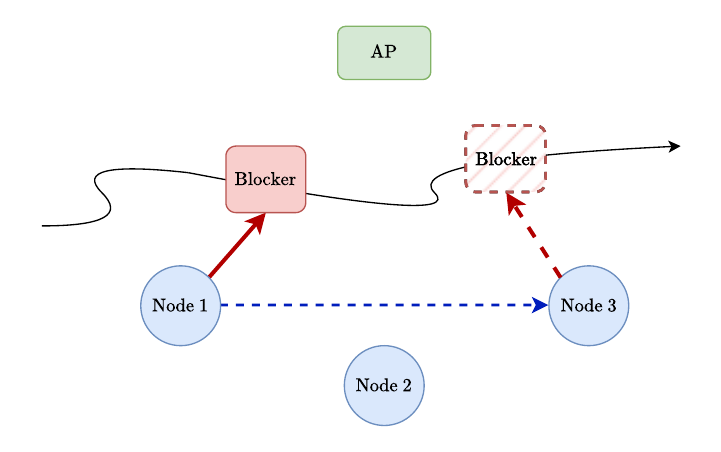}
        \caption{$\tau=t+2$: Node 1 aids Node 3.}
    \end{subfigure}
   \caption{Example scenario illustrating the dependence of node cooperation on the prediction horizon $\tau$. (a) At $t=0$, Node 1 detects a mobile blocker while Nodes 2 and 3 maintain \ac{LOS}. (b) For a short prediction horizon ($\tau=t+1$), Node 2 is informed by Node 1 to predict the incoming blockage. (c) For a longer horizon ($\tau=t+2$), the relevance of information shifts; Node 3 must now cooperate with Node 1 to anticipate the blocker's arrival.}
   \label{fig:example}
\end{figure*}

To overcome these limitations, \emph{\ac{AirComp}} has emerged as a transformative physical-layer technique \cite{10538293}. By integrating computation directly into the transmission phase, 
\ac{AirComp} allows concurrent transmissions to exploit the superposition property of the wireless channel to aggregate signals from multiple nodes over-the-air, enabling all sensors to transmit simultaneously on the same time-frequency resources without decoding individual streams \cite{frey2022over}.
This analog aggregation paradigm is particularly synergistic with \acp{GNN}, where the dominant operation is the aggregation of feature vectors from neighboring nodes \cite{gao2022wide, 10707077, 11073510}. This has given rise to \emph{\acp{AirGNN}} \cite{GaoAirgnns}, architectures that map the graph topology onto the wireless network, utilizing the channel as an intrinsic processing layer of the neural model. 
These architectures jointly optimize communication and learning at the edge by leveraging \ac{AirComp} for the simultaneous transmission of neighboring nodes, effectively performing feature aggregation in the analog domain. 
This grants the system lower communication latency and higher spectral efficiency over traditional orthogonal multiple access schemes, while preserving the accuracy of graph-based learning tasks and enabling scalable distributed inference across edge devices.

In this paper, we investigate this goal-oriented paradigm through the lens of a critical IIoT use case: \emph{\ac{LOS} blockage prediction} for \ac{mmWave} networks. 
High-frequency wireless links are essential for next-generation industrial automation \cite{zhang2025intelligent, shastri2022review}. 
The large bandwidth available at these frequencies supports novel use cases, such as high-resolution wireless sensing, real-time control of mobile robots, and ultra-reliable links for automation \cite{zhou20246, gao2019parallel}. 
However, despite offering wide spectral resources, these frequencies suffer significant molecular absorption, scattering, and blockages, which together  frequently cause severe link degradation and abrupt outages \cite{das2023ambit, chen2023channel}.
Proactive blockage prediction is, therefore, a strict requirement for reliability \cite{yang2024blockage}.
This problem serves as an ideal case study for goal-oriented communications for two reasons. First, it requires ultra-low latency inference that digital schemes struggle to support in dense settings. Second, changing the learning objective (e.g., the prediction horizon) directly affects the level of cooperation required among nodes to fuse information, as illustrated in Fig. \ref{fig:example}, thereby necessitating the joint optimization of communication strategies and learning parameters. By allowing the channel to participate in distributed decision-making via \ac{AirComp}, we do not just aim at efficient radio resource use; instead, we leverage the physics of propagation to solve the prediction task \cite{yang2023environment}.
To this end, we propose \emph{\acp{GO-ST-AirGNN}}, a goal-oriented communication framework based on frequency-selective spatio-temporal \acp{AirGNN}. Our specific contributions are summarized as follows:

\begin{enumerate}

\item We propose \ac{GO-ST-AirGNN}, a unified framework that integrates over-the-air computation with spatio-temporal graph learning. By treating the wireless channel as an analog processing layer, the framework enables scalable, spectral-efficient feature aggregation in dense distributed networks.

\item We develop a goal-oriented \ac{RRM} scheme that acts as a physical attention mechanism. Unlike conventional approaches that maximize throughput, our method optimizes transmission power and spectral resources to minimize downstream prediction error, prioritizing semantically relevant features.

\item We provide theoretical analysis for the spectral efficiency and expressive power of the proposed architecture. We demonstrate that GO-ST-AirGNN achieves spectral-efficient scalability independent of node density by utilizing a constant number of radio resources for feature aggregation. Furthermore, we prove that with sufficient spectral resources the proposed architecture gains the universal approximation capability to emulate standard \acp{MPNN}.

\item We validate the framework using high-fidelity ray-tracing simulations in \ac{IIoT} environments. Results demonstrate that \ac{GO-ST-AirGNN} generalizes effectively to unseen scenarios and domain shifts, matching digital baselines while significantly reducing communication overhead.

\end{enumerate}
The remainder of the paper is organized as follows. Section~\ref{sec:related_work} reviews related works. Section~\ref{sec:system_model} formalizes the system model and the problem formulation. Section~\ref{sec:go_st_airgnn} presents the \ac{GO-ST-AirGNN} architecture.
Section~\ref{sec:theoretical_analysis} formalizes the spectral efficiency and expressive power of \ac{GO-ST-AirGNN}. 
Section~\ref{sec:learning_deployment} discusses the learning procedure and deployment of the proposed architecture. 
Section~\ref{sec:numerical_results} reports the numerical results. Section~\ref{sec:conclusions} concludes the manuscript.

\section{Related Work}
\label{sec:related_work}

Running inference tasks at the edge in next-generation \ac{IIoT} networks requires accurate short-term forecasting of the wireless environment, including proactive blockage prediction \cite{wang2025deep}. Data-driven approaches are emerging for modeling non-linear channel dynamics, often treating the channel as a time-series where historical features reveal latent blocker trajectories \cite{bonfante2021performance}. In this context, \textit{wireless signatures} derived from standard power measurements allow the network to predict blockage events without dedicated hardware \cite{wu2022blockage}.
However, purely \ac{RF}-based methods often lack explicit geometric awareness, necessitating the use of multimodal fusion. While LiDAR-based fusion offers superior spatial precision \cite{wu2022lidar} and RGB-based methods effectively calculate interception probabilities \cite{charan2021vision}, they face cost, privacy, and low-light limitations. Alternatively, radar sensors provide a privacy-preserving option, utilizing range-velocity maps for precise blockage timing \cite{demirhan2022radar}.

Crucially, single-link models fail to capture network-wide spatial correlations. To address this, \acp{ST-GNN} have been adopted to model blockage dependencies by aggregating neighbor information via message passing \cite{SkocajBlocakge}. This spatial efficacy is further validated in power management \cite{jamshidiha2025power}, handover forecasting \cite{mehregan2025gcn}, and dynamic cell-free \ac{MIMO} coverage prediction \cite{jiang2025spatio}, as systematically reviewed in \cite{corradini2025systematic}.
However, standard digital \ac{GNN} implementations relying on orthogonal multiple access suffer from poor scalability, as resource requirements scale linearly with neighborhood size \cite{8870236}. In contrast, \ac{AirComp} mitigates this by exploiting channel superposition for analog aggregation, decoupling bandwidth consumption from the number of neighbors \cite{csahin2023survey}.

\Ac{AirComp} has been extensively studied for \ac{FL}, where the goal is to aggregate local gradients over the air \cite{yang2023one}.
To ensure that signals add up constructively, channel phases must be synchronized. Authors in \cite{wang2023graph} introduce \ac{RIS} to shape the channel environment for AirComp.
The integration of \ac{GNN} with AirComp exploits a natural synergy, as the neighbor aggregation in \ac{GNN} message passing maps directly to the signal superposition property of the wireless channel.
The work in \cite{GaoAirgnns} rigorously analyzes the stability of GNNs when the aggregation is performed over a noisy, fading channel. Instead of training a GNN on ideal data and deploying it on a noisy channel (which leads to degradation), authors incorporate the channel's fading and noise statistics into the training loop, learning neural network weights that are inherently robust to channel impairments. 
The paper in \cite{yang2023implementing} proposes a ``channel-inversion'' transmission strategy: if the physical channel does not match the desired graph topology, nodes can use precoding to transmit their signals such that the effective channel approximates the graph signal, allowing the execution of arbitrary graph convolutions over physical wireless links.
Beyond spectral efficiency, AirGNNs offer privacy advantages. Because individual neighbor features are summed in the air before reaching the receiver, the receiver cannot isolate the raw data of any single neighbor, which provides intrinsic physical-layer privacy for sensitive sensor data \cite{lee2023privacy}.

Goal-oriented and semantic communication have emerged as key enablers for modern control systems and wireless network \cite{feng2024goal, getu2023making}. To realize this approach, 
authors in \cite{zhang2023drl} propose a \ac{DRL}-driven agent that learns to allocate resource blocks based on the semantic importance of the data. Crucially, the reward function here is tied explicitly to task performance metrics 
%, such as classification accuracy, 
rather than conventional data rates. Goal-oriented communication often integrates concepts like the \ac{VoI} alongside \ac{AoI}. As demonstrated in \cite{wu2024goal}, minimizing \ac{AoI} is not always optimal; an older packet with high semantic density (e.g., a critical keyframe) may yield a higher reward than fresher, redundant updates.
Finally, this focus on utility naturally leads to semantic compression, where the objective shifts from transmitting data efficiently to transmitting less data overall \cite{cheng2023resource}. %Techniques such as those exploring rate-splitting multiple access (RSMA) for semantic transmission propose filtering features at the transmitter side \cite{cheng2023resource}.

Despite these advances, a unified framework integrating the spatial awareness of \acp{ST-GNN}, the scalability of \ac{AirComp}, and goal-oriented utility is missing. Existing AirGNN frameworks typically treat wireless transmission and graph inference as distinct optimization stages, limiting efficacy in dynamic blockage prediction. To address this, we propose \ac{GO-ST-AirGNN}, a framework that jointly optimizes \ac{RRM} and learning parameters, ensuring resource allocation is driven explicitly by the semantic value of the prediction task.

\section{\ac{LOS} Blockage Prediction Problem}
\label{sec:system_model}

\begin{figure}[t]
    \centering
    \includegraphics[trim= {0 0 0 0}, clip, width=0.93\columnwidth]{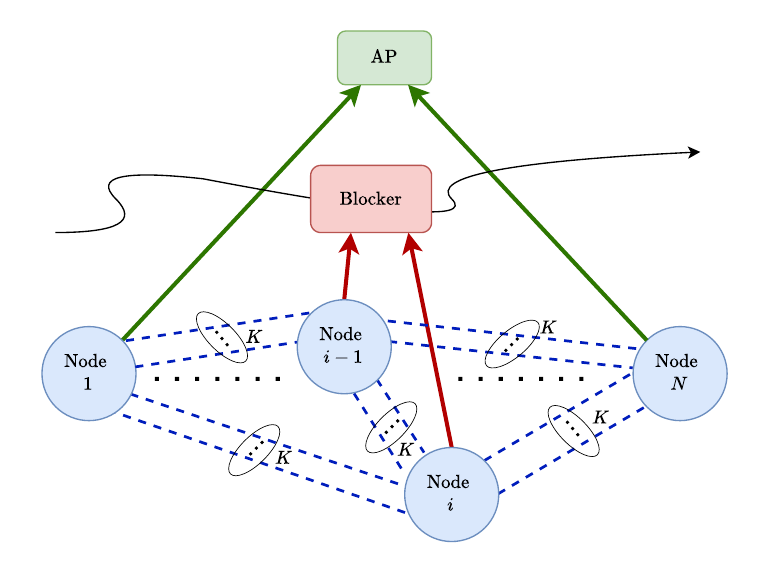}
    \caption{The network contains %consists of 
    a set of wireless nodes $\mathcal{N}$ served by a central \ac{AP}. The direct links are susceptible to blockage by dynamic obstacles, leading to \ac{NLOS} conditions. To anticipate these events, nodes cooperate by exchanging local features $\mathbf{x}_{i,t}$ over $K$ parallel dynamic communication graphs $\{\mathcal{G}_{t,k}\}_{k=1}^K$, leveraging \ac{AirComp} to aggregate information, and predicting the future blockage status $y_{i, t+\tau}$.}
    \label{fig:sys_model}
\end{figure}

\subsection{System Model}
We consider a wireless \ac{IIoT} network comprising a set of full-duplex single-antenna wireless nodes $\mathcal{N} = \{1, \dots, N\}$ served by a central controller or \ac{AP}. The nodes operate in the mmWave frequency band and necessitate strict \ac{LOS} connectivity to the \ac{AP}.
We consider a dynamic system where time is slotted into discrete intervals $t \in \{1, \dots, T\}$. At any time slot $t$, the propagation environment may vary due to the movement of mobile obstacles within the industrial area. These obstacles can temporarily shadow the communication links, causing \ac{NLOS} blockage events.
The critical link requiring monitoring is the direct channel between node $i \in \mathcal N$ and the \ac{AP}, denoted by $\mathbf h_{i,t}^{\text{AP}} \in \mathbb{C}$.
To anticipate the reliability of this link, we define a binary blockage label $y_{i, t+\tau} \in \{0, 1\}$ with %where $\tau$ is %for 
a future prediction horizon $\tau$. Here, $y_{i, t+\tau}=1$ indicates a blockage (\ac{NLOS}) event, while $y_{i, t+\tau}=0$ denotes a clear \ac{LOS} condition.

Since nodes do not possess global knowledge of the environment, they rely on local channel estimations to infer future states. At each time slot $t$, each node $i$ constructs a feature vector $\mathbf{x}_{i,t} \in \mathbb{R}^L$ encapsulating the local temporal evolution of the wireless channel. For instance, $\mathbf{x}_{i,t}$ may consist of a sliding window of historical channel gains as
\begin{equation}\label{eq:localFeatures}
    \mathbf{x}_{i,t} = [|\mathbf h_{i,t}^{\text{AP}}|, |\mathbf h_{i,t-1}^{\text{AP}}|, \dots, |\mathbf h_{i,t-L+1}^{\text{AP}}|]^\top.
\end{equation}
To cooperate and improve prediction accuracy, nodes can exchange their local features $\mathbf{x}_{i,t}$ at time slot $t$ over $K$ orthogonal frequency resources (or subcarriers), modeled as $K$ parallel dynamic communication graphs $\{\mathcal{G}_{t,k}\}_{k=1}^K$. %, corresponding to.
The existence of edges in these graphs is determined by the wireless channel conditions and the resource allocation strategy. Specifically, at each time $t$, let $\mathbf h_{i,j,k,t} \in \mathbb{C}$ be the channel coefficient between node $i$ and node $j$ on frequency $k$. 
Every node $i$ allocates a transmit power vector $\mathbf{p}_{i,t} = [p_{i,1,t}, \dots, p_{i,K,t}]^\top \in \mathbb{R}_+^K$ across the $K$ subcarriers, and 
the edge set for the $k$-th frequency is defined as 
\begin{equation}
\label{eq:edge_set_condition}
\Edges_{t,k}
= \Big\{ (i,j)\in\mathcal{N}^2 : i\neq j,\;
\frac{p_{i,k,t}\,\mathbb{E}\!\left[|\mathbf h_{i,j,k,t}|^2\right]}{\sigma^2}
\ge \gamma_{\min}
\Big\}.
\end{equation}
where the channel expectation $\mathbb{E}\!\left[|\mathbf h_{i,j,k,t}|\right]^2$ is computed over small scale fading, $\sigma^2$ is the thermal noise at the receiver, and $\gamma_{\min}$ is the receiver sensitivity. 
Consequently, the topology of the $k$-th graph $\mathcal{G}_{t,k}$ is dynamic and frequency-selective; an edge $(i,j)$ exists only if condition \eqref{eq:edge_set_condition} is satisfied.
An overview of the scenario is provided in Fig. \ref{fig:sys_model}.

\subsection{Problem Formulation}
The primary objective of the network is to predict the future blockage status of all nodes. We formulate this as a supervised learning task where the predicted status $\hat{y}_{i, t+\tau}$ is the output of a function $f_i(\cdot\,; \boldsymbol \theta)$ parameterized by $\boldsymbol \theta$.
The prediction function relies on the 
feature states of the nodes $\mathbf{X}_t = [\mathbf x_{1,t}, \ldots,\mathbf x_{N,t}]$ and the communication graphs shaped by the power allocation 
$\mathbf{p}_t = [\mathbf{p}_{1,t}, \ldots, \mathbf{p}_{N,t}]$ [cf. \eqref{eq:edge_set_condition}] as
\begin{equation}
    \hat{y}_{i, t+\tau} = f_i \left( \mathbf{X}_t, \left\{ \mathcal{G}_{t,k}(\mathbf{p}_t) \right\}_{k=1}^K\,; \boldsymbol \theta \right).
\end{equation}
To measure the prediction quality, we utilize the \ac{BCE} loss function. The network-wide loss at time $t$ is defined as:
\begin{equation}
\begin{split}
    \mathcal{J}_t(\boldsymbol{\theta}, \mathbf{p}_{t}) = -\frac{1}{N} \sum_{i=1}^{N} \mathbb{E} \Big[ &y_{i, t+\tau} \log\big(\hat{y}_{i, t+\tau}\big) + \\
    &(1 - y_{i, t+\tau}) \log\big(1 - \hat{y}_{i, t+\tau}\big) \Big].
\end{split}
\end{equation}
We aim to jointly optimize the prediction model parameters $\boldsymbol{\theta}$ and the goal-oriented power allocation $\mathbf{p}_t$. Unlike standard \ac{RRM} which maximizes data rates, our goal-oriented framework designs 
the power allocation strategy 
to shape the 
topology of communication graphs $\mathcal{G}_{t,k}(\mathbf{p}_t)$ and 
support the prediction model to minimize the prediction error. Given a 
total power budget $P_{\text{tot}}$ at each node, the joint optimization problem is formulated as:

\begin{subequations}
\begin{align}
    \min_{\boldsymbol{\theta}, \mathbf{p}_{t}} \quad & \mathcal{J}_t(\boldsymbol{\theta}, \mathbf{p}_{t}) \\
    \text{s.t.} \quad & \sum_{k=1}^{K} p_{i,k,t} \leq P_{\text{tot}}, \quad \forall i \in \mathcal{N}, \label{eq:power_constraint1}\\
    & p_{i,k,t} \geq 0, \quad \forall i \in \mathcal{N}, \forall k \in \{1,\dots,K\}. \label{eq:power_constraint2}
\end{align}
\end{subequations}
This formulation encourages a goal-oriented communication strategy: by allocating transmit power to specific subcarriers, nodes effectively establish edges in the dynamic graphs $\{\mathcal{G}_{t,k}\}_{k=1}^K$, enabling the flow of features $\mathbf{x}_{i,t}$ required for cooperative sensing.

\section{Goal-Oriented Architecture}
\label{sec:go_st_airgnn}

\begin{figure*}[t]
    \centering
    \includegraphics[trim= {0 0 0 0}, clip, width=0.93\textwidth]{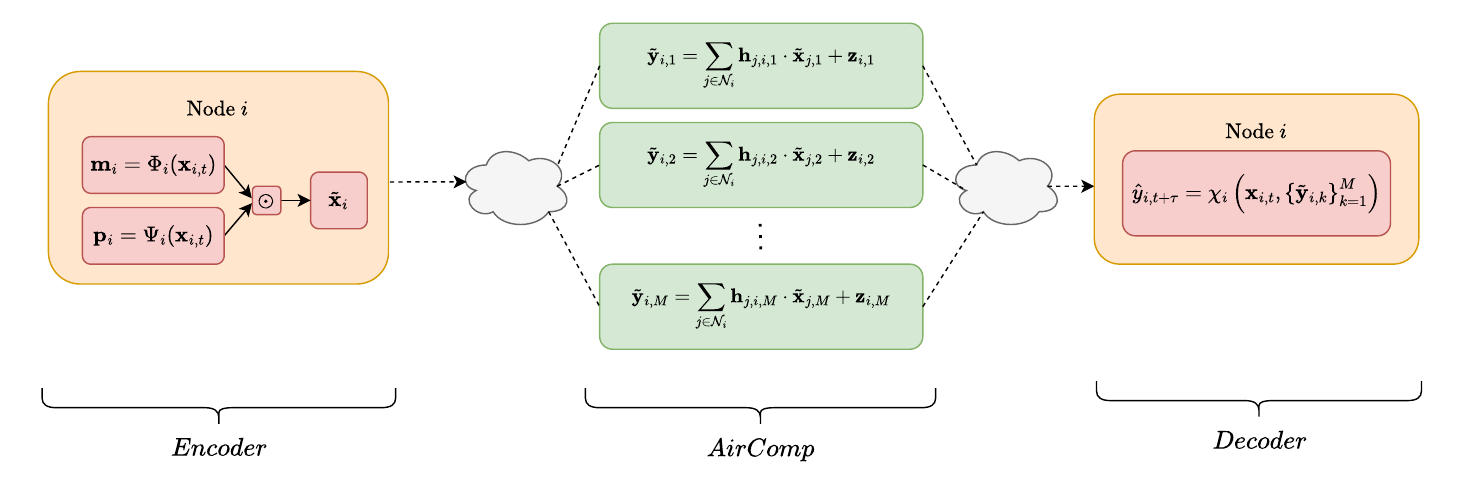}
    \caption{Overview of the proposed \ac{GO-ST-AirGNN} architecture. 
(Left) At the transmitter, node $i$ uses a semantic encoder $\Phi_i$ and a resource manager $\Psi_i$ to jointly generate the message embedding $\mathbf{m}_i$ and the goal-oriented power allocation vector $\mathbf{p}_i$. 
(Center) The wireless channel acts as a computation layer (\ac{AirComp}), naturally aggregating neighbor signals via electromagnetic superposition across the parallel dynamic communication graphs. (Right) At the receiver, the semantic decoder $\chi_i$ processes the aggregated analog signals $\tilde{\mathbf{y}}_{i,k}$ together with the local features to predict the future blockage status $\hat{y}_{i, t+\tau}$.}
    \label{fig:methodology}
\end{figure*}

To address the joint problem of resource allocation and blockage prediction, we propose \ac{GO-ST-AirGNN}, a unified goal-oriented framework that couples over-the-air computation with spatio-temporal graph learning. Unlike conventional \acp{GNN} that assume perfect digital exchanges of node features followed by local processing, the proposed architecture integrates the communication channel directly into the learning loop, treating the physical medium as an inherent processing layer.
This design leverages the waveform superposition property of the wireless channel to perform spectral-efficient feature aggregation via \ac{AirComp}. Moreover, it accounts for both spatial and temporal information for resource allocation and blockage prediction, which addresses the joint problem in a comprehensive manner. Specifically, 
the \ac{GO-ST-AirGNN} framework is built upon three fundamental pillars:

\begin{itemize}
    \item \textit{Over-the-air graph aggregation}. Instead of digitizing, packetizing, and scheduling feature vectors, nodes transmit analog signals simultaneously. \ac{GO-ST-AirGNN} leverages the superposition property of the wireless channel to naturally compute convolutions over graphs fundamental to \acp{GNN}.
    
    \item \textit{Goal-oriented communications}.
    By optimizing the resource allocation vector $\mathbf{p}_t$, \ac{GO-ST-AirGNN} dynamically weighs the edges of the communication graph and optimizes the topology to support the blockage prediction model and minimize the prediction loss. This goal-oriented RRM introduces a physical graph attention mechanism. Different from classical graph attention that is typically computed at the receiver, it applies attention directly at the transmitter.
    
    \item \textit{Spatio-temporal reasoning}. Blockage events in industrial environments are rarely isolated (e.g., they result from continuously moving obstacles). In this context, \ac{GO-ST-AirGNN} processes temporal sequences of local features $\mathbf{X}_t$ over the graph temporal process $\{\mathcal{G}_{t,k}\}$ to capture the spatio-temporal correlations necessary for future blockage prediction.
\end{itemize}
The architecture decomposes the blockage prediction task into three stages: (i) a \emph{semantic encoder}, which determines the transmission policy of local features, (ii) an \emph{over-the-air aggregation layer}, which conducts the feature aggregation with %includes an 
\ac{AirComp}, %step, 
and (iii) a \emph{semantic decoder}, which performs the downstream blockage prediction. The overall architecture is depicted in Fig.~\ref{fig:methodology}.

\subsection{Semantic Encoder and Transmission Policy}\label{subsec:Encoder}

In the proposed \ac{GO-ST-AirGNN} architecture, the transmission stage at each node $i$ is decoupled into two parallel learnable processes: \textit{semantic embedding} and \textit{resource management}. The former learns higher-level representations from local features to determine transmission content, while the latter learns the frequency-aware power allocation strategy to determine transmission policy. 
At time slot $t$, node $i$ processes its local feature vector $\mathbf{x}_{i,t}$ through two neural networks:

\subsubsection{Message Embedding}
First, a semantic encoder $\Phi_i(\cdot)$, parameterized by weights $\boldsymbol{\theta}_{i, \Phi}$, maps the input features to a normalized message embedding vector 
\begin{equation}
    \mathbf{m}_{i,t} = \frac{\Phi_i(\mathbf{x}_{i,t}; \boldsymbol{\theta}_{i, \Phi})}{|\Phi_i(\mathbf{x}_{i,t}; \boldsymbol{\theta}_{i, \Phi})|^2}. 
\end{equation} 
This vector represents the semantic content to be communicated across the $K$ subcarriers. %through graph convolution operations.
The normalization ensures that $\mathbf{m}_{i,t}$ has unit power prior to being scaled by the transmit power allocation, specified in the subsequent section. %graph convolution operations.

\subsubsection{Resource Management}
Second, %Simultaneously, 
a resource management module $\Psi_i(\cdot)$, parameterized by weights $\boldsymbol{\theta}_{i, \Psi}$, maps the input features %generates 
to a power allocation vector 
\begin{equation}
    \mathbf{p}_{i,t} = \Psi_i(\mathbf{x}_{i,t}; \boldsymbol{\theta}_{i, \Psi}).
\end{equation}
This vector determines the transmission amplitude for each subcarrier. The output of $\Psi_i$ is constrained to be non-negative and to satisfy \eqref{eq:power_constraint1} and \eqref{eq:power_constraint2} through a softmax-based activation function at the output layer.

\subsubsection{Goal-Oriented Transmission}
Finally, %The final 
the transmitted signal $\tilde{\mathbf{x}}_{i,t}$ is obtained via Hadamard product of the power allocation %resource 
vector and the message embedding, i.e., %, that is 
\begin{equation}\label{eq:anaSignal}
    \tilde{\mathbf{x}}_{i,t} = \mathbf{p}_{i,t} \odot \mathbf{m}_{i,t}.
\end{equation}
In this context, the semantic encoder not only learns the transmitted content but also learns the transmission power; together, yielding a goal-oriented communication strategy. %a goal-oriented resource management strategy. 
By optimizing $\boldsymbol{\theta}_{i, \Phi} \text{ and } \boldsymbol{\theta}_{i, \Psi}$ w.r.t. the final prediction loss, the node weighs message embeddings with physical attention by allocating power to subcarriers and formulates communication graphs to emphasize features that carry the most relevant information regarding the blockage event.

\subsection{Over-the-Air Aggregation Layer}
The core component of \ac{GO-ST-AirGNN} is the use of the wireless channel as a computational layer to perform graph convolutions with over-the-air computation (AirComp). Specifically, it employs an \ac{AirComp} scheme to aggregate the generated message embeddings [Section \ref{subsec:Encoder}] from neighbors via uncoded analog communication \cite{csahin2023survey}. Unlike digital schemes that transmit data packets orthogonally, \ac{AirComp} leverages the superposition property of the wireless channel to perform data aggregation directly in the analog domain. This grants superior spectral efficiency and ultra-low latency, enabling the required communication resources to remain constant regardless of the number of transmitting neighbors. 

Each node $i$ transmits $\tilde{\mathbf{x}}_{i,t}$ across the $K$ available frequency subcarriers. We denote by $\tilde{\mathbf{x}}_{i,k,t}$ the analog signal transmitted by node $i$ on subcarrier $k$. 
The signal received at a generic node $j$ is the electromagnetic superposition of transmitted signals from its neighbors.
It is important to note that aggregation of neighboring signals occurs spatially but is maintained orthogonally across frequencies\footnote{To operate the $K$ parallel dynamic graphs, the total bandwidth $B$ is divided into $K$ uniformly spaced subcarriers. To strictly avoid inter-carrier interference, each subcarrier $k$ is allocated a bandwidth of $B_k = \tfrac{B-\Delta_f}{K}$, separated by a guard-band $\Delta_f$.}. 
Consequently, the total bandwidth is utilized such that the signals transmitted on different subcarriers remain spectrally distinct.  
Specifically, the signal received at node $i$ on frequency $k$ is given by:
\begin{equation}
    \tilde{\mathbf y}_{i,k,t} = \sum_{j \in \mathcal{N}_i} \mathbf h_{j,i,k,t}\ \tilde{\mathbf{x}}_{j,k,t} + \mathbf z_{i,k,t},
    \label{eq:ota_reception}
\end{equation}
where $\mathbf z_{i,k,t} \sim \mathcal{CN}(0, \sigma^2)$ represents the additive white Gaussian noise and $\sigma^2$ is the noise variance. 
The node $i$ then collects the vector of received signals $\tilde{\mathbf{y}}_{i, t} = [\tilde{\mathbf{y}}_{i,1,t}, \dots, \tilde{\mathbf{y}}_{i,K,t}]$ over the $K$ subcarriers. This is %effectively receiving 
a $K$-dimensional aggregated message, where %with 
each dimension corresponds to a specific communication graph $\G_{t,k}$ and the graph topology is determined by the %respective 
allocated power.

\subsection{Semantic Decoder and \ac{LOS} Blockage Prediction}

The final stage of the \ac{GO-ST-AirGNN} pipeline takes place at the receiver side of node $i$. Here, the physical analog signals $\tilde{\mathbf{y}}_{i, t}$ received over-the-air are processed to predict future \ac{LOS} blockages.
To recover the specific blockage probability, node $i$ employs a semantic decoder function $\chi_i(\cdot)$, parameterized by weights $\boldsymbol{\theta}_{i, \chi}$. This module acts as the readout head of the \ac{GNN}.
Since blockage events are dynamic processes, the decoder must reason over time. Therefore, we define the input to the decoder not just as the instantaneous reception, but as a history of both local features and spatially aggregated context.
Specifically, we denote by $\mathbf{s}_{i,t}$ the combined state at time $t$:
\begin{equation}\label{eq:combinedState}
    \mathbf{s}_{i,t} = \left[ \mathbf{x}_{i,t} \, \| \, \mathbf{y}_{i, t} \right],
\end{equation}
where $\cdot\|\cdot$ denotes concatenation.
The semantic decoder $\chi_i$ processes a sliding window of these states to capture temporal dependencies. The estimated blockage probability $\hat{y}_{i, t+\tau}$ is given by:
\begin{equation}\label{eq:predictionProb}
    \hat{y}_{i, t+\tau} = \chi_{i} \left( \left\{ \mathbf{s}_{i, t'}, \right\}_{t'=t-L+1}^{t} ; \boldsymbol{\theta}_{i, \chi} \right).
\end{equation}
Functionally, $\chi_i$ can be implemented as a sequence-based model (e.g., \acp{RNN} or a temporal convolutional layer) followed by a binary classification head. We can then make predictions based on $\hat{y}_{i, t+\tau}$.

\section{Theoretical Analysis}
\label{sec:theoretical_analysis}

\subsection{Spectral Efficiency}\label{subsec:efficiency}

A critical challenge in dense \ac{IIoT} deployments is maintaining spectral-efficient operations as the network density increases. In this section, we formalize the advantage of \ac{GO-ST-AirGNN} in terms of communication scalability, compared to standard digital baselines.
We consider a network modeled by a graph $\mathcal{G}=(\mathcal{N}, \mathcal{E})$, where each node $i \in \mathcal{N}$ must aggregate features from its neighborhood $\mathcal{D}_i$ over $K$ orthogonal resource blocks. We define the \textit{communication cost} $T_{\text{air}}$ and $T_{\text{dig}}$ as the number of time slots required for all nodes in $\mathcal{N}$ to complete a graph convolution (aggregation) step across the network for analog and digital approaches, respectively.

\textbf{Proposition 1 ($O(1)$ Communication Cost in \ac{GO-ST-AirGNN}).}
Let $\Delta(\mathcal{G}) = \max_{i \in \mathcal{N}} |\mathcal{D}_i|$ denote the maximum node degree, which represents the size of the largest neighborhood in $\mathcal{G}$, and, therefore, the number of messages that the node with most neighbors receives during an aggregation step.
The communication cost for \ac{GO-ST-AirGNN} ($T_{\text{air}}$) and the lower bound for a digital baseline ($T_{\text{dig}}$) scale as follows:
\begin{align}
    T_{\text{air}} = 1, \quad
    T_{\text{dig}} \ge \left\lceil \frac{\Delta(\mathcal{G})}{K} \right\rceil,
\end{align}
where $\lceil\cdot\rceil$ is the ceil operator.

\textit{Proof.}
See the Appendix.

Proposition 1 states that the asymptotic communication complexity of \ac{GO-ST-AirGNN} is $O(1)$, whereas that of the digital scheme is $O(\Delta(\mathcal{G}))$. Crucially, the former is independent of the graph connectivity, while the latter is degree-limited. This result establishes a critical scalability advantage of our framework. As the network gets denser, the proposed \ac{GO-ST-AirGNN} maintains constant communication cost $T_{\text{air}}$, independently of the local node density. In contrast, the spectral-efficiency of digital approaches degrades, creating a bottleneck for real-time inference. 

\subsection{Expressive Power}
\label{sec:expressivity}

While the primary advantage of \ac{GO-ST-AirGNN} lies in its spectral efficiency and low-latency analog aggregation via \ac{AirComp}, it is theoretically significant to establish its representational capacity compared to standard digital \acp{GNN}. 
Typically, \acp{GNN} rely on the \ac{MPNN} framework, where nodes exchange discrete feature vectors to compute updates.
% A concern with analog aggregation is that the physical channel forces a specific aggregation operator (i.e., summation), potentially limiting the model's ability to learn complex non-linear interactions (e.g., max-pooling or attention mechanisms) that digital \acp{MPNN} can implement seamlessly.
A concern with analog aggregation may be that the physical channel forces a specific aggregation operator (i.e., summation) and restricts the degrees of freedom to model graph structures, as the aggregation is bound by channel conditions and \ac{RRM} rather than allowing for the full signal separation typical of digital \acp{MPNN}. 
In this section, we formally analyze the expressive power of the proposed architecture at a fixed time slot $t$. 
We demonstrate that with sufficient resource units, i.e.,  
if the number of subcarriers $K$ is sufficiently large %large enough 
to ensure every neighbor $j$ of a target node $i$ transmits on a unique, interference-free frequency, GO-ST-AirGNN can decode individual messages. Consequently, it satisfies the prerequisites to emulate standard MPNN neighborhood aggregation and is capable of approximating any standard MPNN layer at any time slot $t$.
Specifically, in a generic update layer of a standard \ac{MPNN}, for a node $i$ with hidden state $\mathbf{u}_i \in \mathbb{R}^d$, the updated state $\mathbf{u}_i'$ is computed as
\begin{equation}
    \mathbf{u}_i' = \mathcal{U} \left( \mathbf{u}_i, \bigoplus_{j \in \mathcal{N}_i} \mathcal{M}(\mathbf{u}_i, \mathbf{u}_j) \right),
    \label{eq:generic_mpnn}
\end{equation}
where $\mathcal{M}(\cdot)$ is a learnable message function, $\bigoplus$ is a permutation-invariant aggregation operator (e.g., sum, mean, or max), and $\mathcal{U}(\cdot)$ is a state update function. In this digital abstraction, messages are transmitted reliably and orthogonally.
We now show that the proposed analog architecture subsumes the generic \ac{MPNN} formulation under specific conditions.

\textbf{Proposition 2 (Universal Approximation via GO-ST-AirGNN).}
If the number of subcarriers is $K = K'$, where $K'$ is the number of subcarriers 
required to ensure that for every node $i$, all incoming signals from its neighbors $j \in \mathcal N_i$ are mapped to mutually orthogonal frequencies, 
and assuming a high-\ac{SNR} regime (i.e., $\frac{P_{\text{tot}}}{\sigma^2} \to \infty$), the \ac{GO-ST-AirGNN} architecture can approximate the generic \ac{MPNN} update rule defined in \eqref{eq:generic_mpnn} with arbitrary precision at any given time slot $t$.

\textit{Proof.}
See the Appendix.

Proposition 2 establishes that the analog nature of \ac{GO-ST-AirGNN} does not inherently limit its reasoning capabilities compared to digital counterparts. It proves that, with sufficient bandwidth, the system gains the degrees of freedom to emulate any arbitrary message passing logic via over-the-air aggregation. 
It is interesting to see that the number of subcarriers %Practically, 
$K$ regulates the trade-off between expressivity and latency.
Approaching $K \approx K'$ recovers the full separation capabilities of digital \acp{MPNN} at the cost of bandwidth, whereas a smaller $K$ fully exploits \ac{AirComp} to maximize spectral efficiency at the expense of signal resolution.
Moreover, in moderate SNR regimes, the stochasticity due to thermal noise during the over-the-air aggregation step can act as a regularization term and improve robustness against overfitting.

\section{Learning Procedure and Deployment}
\label{sec:learning_deployment}

The proposed \ac{GO-ST-AirGNN} integrates communication and computation into a single differentiable pipeline. This structure allows us to adopt a \ac{CTDE} paradigm \cite{gao2023decentralized, ctde2, ctde1}. 
While the analog aggregation over physical channels is employed in both training and execution, %phases, 
the training phase leverages centralized parameter sharing to jointly optimize transmission and prediction rules. In the deployment phase, nodes decouple from this central coordination to execute the learned policies in a fully distributed manner using only local information.

\subsection{End-to-End Optimization}
The training objective is to find the optimal set of parameters $\boldsymbol{\Theta} = \{ \boldsymbol{\theta}_{i,\Phi}, \boldsymbol{\theta}_{i,\Psi}, \boldsymbol{\theta}_{i,\chi} \}_{i=1}^N$ that minimizes the long-term blockage prediction loss.
Since %Crucially, because 
the wireless channel operations in \ac{AirComp} are linear and differentiable, the entire system from the semantic encoder at the transmitter to the semantic decoder at the receiver can be modeled as a single computational graph.
We consider a training dataset $\mathcal{T}$ consisting of sequences of channel realizations, node features, and ground-truth blockage labels. The optimization problem is to minimize the expected loss over this dataset:
\begin{equation}
    \min_{\boldsymbol{\Theta}} \mathbb{E}_{\mathcal{T}} \left[ \mathcal{J}(\boldsymbol{\Theta}) \right].
\end{equation}
To solve this, we employ a stochastic gradient descent approach.  
As training is centralized, this error signal propagates from the receiver, through the channel layer, and back to the transmitters, allowing the semantic encoders and decoders to jointly adapt to the wireless environment.

\textit{Remark on the \ac{CTDE} Paradigm}. The proposed \ac{CTDE} framework assumes that parameter optimization is performed jointly at a central location, requiring model parameters exchanges over the network. Consequently, we envision the training phase occurring \textit{offline} within a high-fidelity digital twin environment. For future work, we consider extending the framework to a \textit{distributed training-distributed execution} setting.

\subsection{Decentralized Execution}
Once the training converges, the optimized parameter sets $\{ \boldsymbol{\theta}_{i, \Phi}, \boldsymbol{\theta}_{i, \Psi}, \boldsymbol{\theta}_{i, \chi} \}_{i=1}^N$ are deployed to the respective nodes.
During the online inference stage, the system operates without any central coordination or global channel knowledge. The execution at each node $i$ is strictly local:
\begin{enumerate}
    \item \textit{Local sensing}. Node $i$ collects its local features %measurements 
    $\mathbf{x}_{i,t}$ [cf. \eqref{eq:localFeatures}].
    \item \textit{Goal-oriented transmission}. With its trained %Using its frozen 
    local models $\Phi_i$ and $\Psi_i$, node $i$ %the node 
    computes the transmitted content $\mathbf{m}_{i,t}$ and the transmission power $\mathbf{p}_{i,t}$ based on $\mathbf{x}_{i,t}$, and transmits the analog signal $\tilde{\mathbf{x}}_{i,t}$ [cf. \eqref{eq:anaSignal}]. 
    \item \textit{Over-the-air aggregation}. Node $i$ aggregates its neighbors' signals through physical channels %the physical channel naturally aggregates the neighbor signals 
    via \ac{AirComp} [cf. \eqref{eq:ota_reception}].
    \item \textit{Local blockage prediction}. Node $i$ concatenates the combined state $\textbf{s}_{i,t}$ [cf. \eqref{eq:combinedState}], %receives $\tilde{\mathbf{y}}_{i,t}$ 
    and processes it via $\chi_i$ to compute the local blockage probability and make the prediction [cf. \eqref{eq:predictionProb}]. %the local blockage probability.
\end{enumerate}
This decentralized architecture ensures that the computational complexity at each node remains constant regardless of the network size, enabling scalable, real-time inference.

\textit{Remark on Synchronization}. Coherent \ac{AirComp} requires strict phase and time synchronization among transmitting nodes to ensure constructive signal superposition \cite{csahin2023survey}. To this end, many approaches have been investigated. Standard techniques such as timing advance (TA) in 4G Long Term Evolution can facilitate synchronization \cite{mahmood2019time} or reference signal broadcasting can be employed to align transmission windows \cite{abari2015airshare}. Moreover, recent literature addresses synchronization challenges by either exploiting signal approximations to bypass precise timing \cite{csahin2022over}, or by adopting hierarchical architectures with multi-phase relaying \cite{wang2022amplify}.
Finally, the propagation characteristics of \ac{mmWave} frequencies inherently induce highly sparse graph topologies $\mathcal{G}_{t,k}$. Unlike sub-6 GHz networks, 
%where omnidirectional interference necessitates global synchronization, 
the strict directionality and blockage of high-frequency links isolate interactions to small subsets of neighbors. This spatial sparsity effectively relaxes the synchronization requirement from a global constraint to a feasible set of local pairwise alignments \cite{chen2018over}.

\section{Numerical Results}
\label{sec:numerical_results}

In this section, we evaluate the performance of the proposed \ac{GO-ST-AirGNN} framework using a high-fidelity digital twin of the BI-REX Competence Center\footnote{See: \url{https://bi-rex.it/en/}.}, simulated via NVIDIA Sionna RT \cite{sionna_ref}. The visual representation of the simulation environment is provided in Fig.~\ref{fig:sionna_birex}.

\subsection{Experimental Setup}

\subsubsection{Simulation Environment and Mobility}
The physical propagation environment is modeled as an indoor industrial factory floor (approx. $300$ m$^2$) populated with static metallic machinery and dynamic obstacles.
The primary source of channel variability and blockage is an \ac{AMR} moving through the factory. The \ac{AMR} follows randomized trajectories with a typical speed of $1\,\text{m/s}$.
We discretize time into steps of $\Delta t \approx 300\,\text{ms}$. Each simulation sequence consists of $T=50$ time steps, corresponding to approximately $15\,\text{m}$ of robot travel path per episode. This high temporal resolution captures the gradual onset and resolution of shadowing events caused by the moving blocker.

\subsubsection{Network Configuration and Frequency Planning}
\begin{figure}[t]
    \centering
    \includegraphics[width=0.85\linewidth]{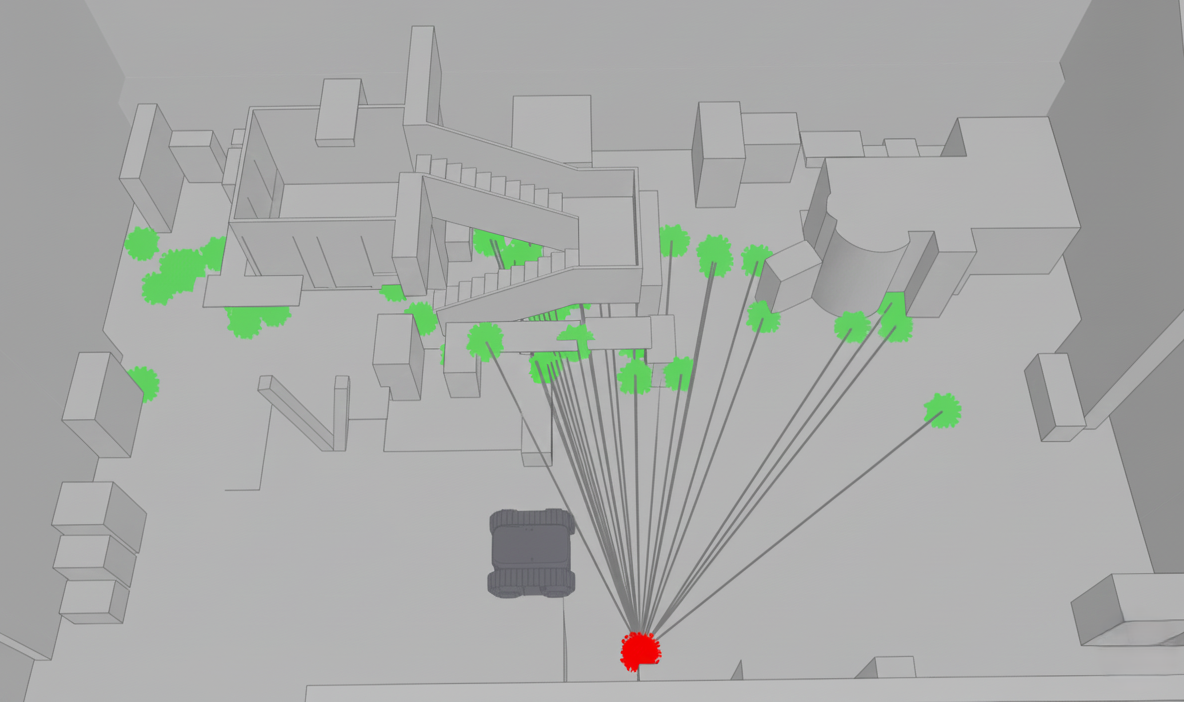} 
    \caption{Digital twin of the BI-REX Competence Center simulated in NVIDIA Sionna RT. The network consists of distributed \ac{IIoT} nodes (green spheres) served by a central \ac{AP} (red sphere). A mobile robot acts as a dynamic blocker, traversing the factory floor and temporarily obstructing the direct \ac{LOS} communication links between devices and \ac{AP}.}
    \label{fig:sionna_birex} 
\end{figure}

The network consists of a set of static \ac{IIoT} nodes (green nodes in Fig.~\ref{fig:sionna_birex}) communicating with a central \ac{AP} (red node).
The communication operates in the mmWave band, centered at $f_c = 28\,\text{GHz}$. To enable the parallel graph topologies, we allocate $K$ orthogonal subcarriers. We explore configurations with $K \in \{1, 2, 4, 8, 16\}$.
Each subcarrier $k$ is assigned a bandwidth of $B_k = 400\,\text{kHz}$, separated by a guard band $\Delta_f = 100\,\text{kHz}$ to prevent inter-carrier interference (e.g., $f_1 = 28.0000\,\text{GHz}$, $f_2 = 28.0005\,\text{GHz}$, etc.).
The nodes operate under a strict power constraint of $P_{\text{tot}} = 0\,\text{dBm}$, which must be dynamically allocated across the $K$ active subcarriers.

\subsubsection{Dataset Generation and Splitting}
We generate a training set $\mathcal{T}_\text{train}$ and a test set $\mathcal{T}_\text{test}$, each comprising $1000$ independent time-series realizations with $N=25$ nodes.
For each realization, the \ac{IIoT} nodes are randomly deployed in valid positions within the facility, ensuring spatial diversity. The two datasets include disjoint node positions to evaluate the inductive generalization capability of our framework %the model 
to unseen topologies.

\begin{figure*}[t]
    \centering
    \begin{subfigure}{0.45\textwidth}
        \includegraphics[width=\linewidth]{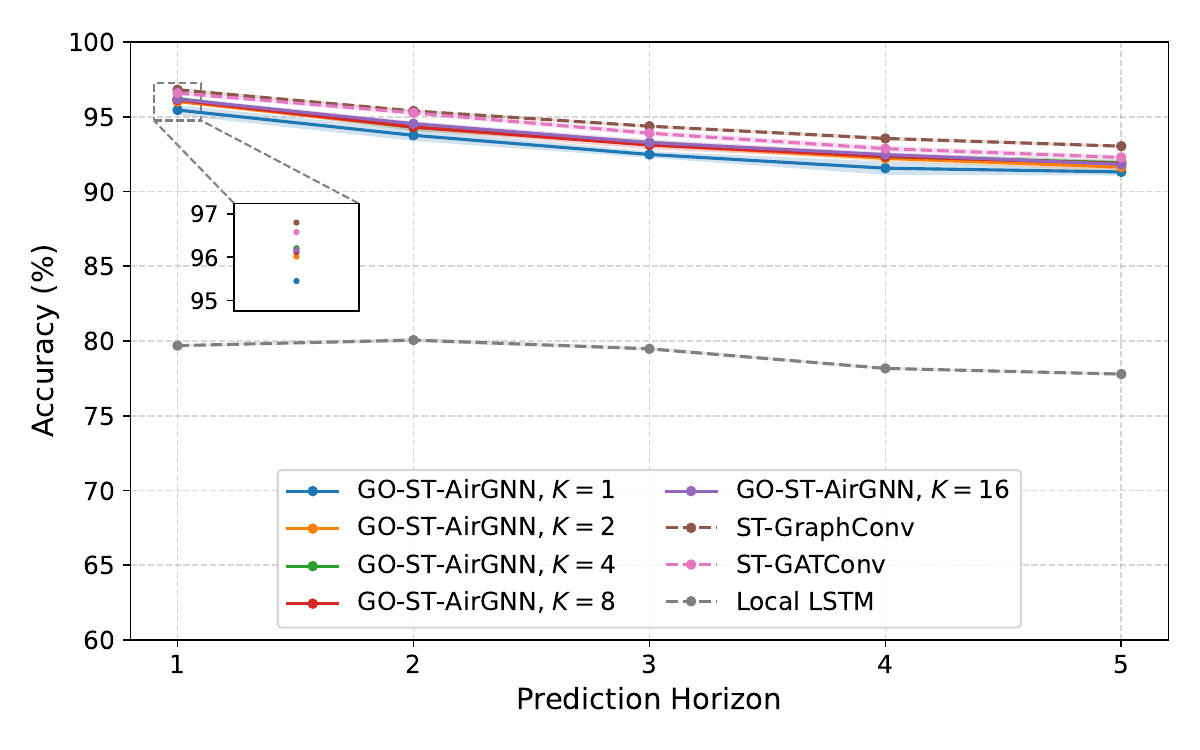}
        \caption{Accuracy.}
    \end{subfigure}
    \quad \quad
    \begin{subfigure}{0.45\textwidth}
        \includegraphics[width=\linewidth]{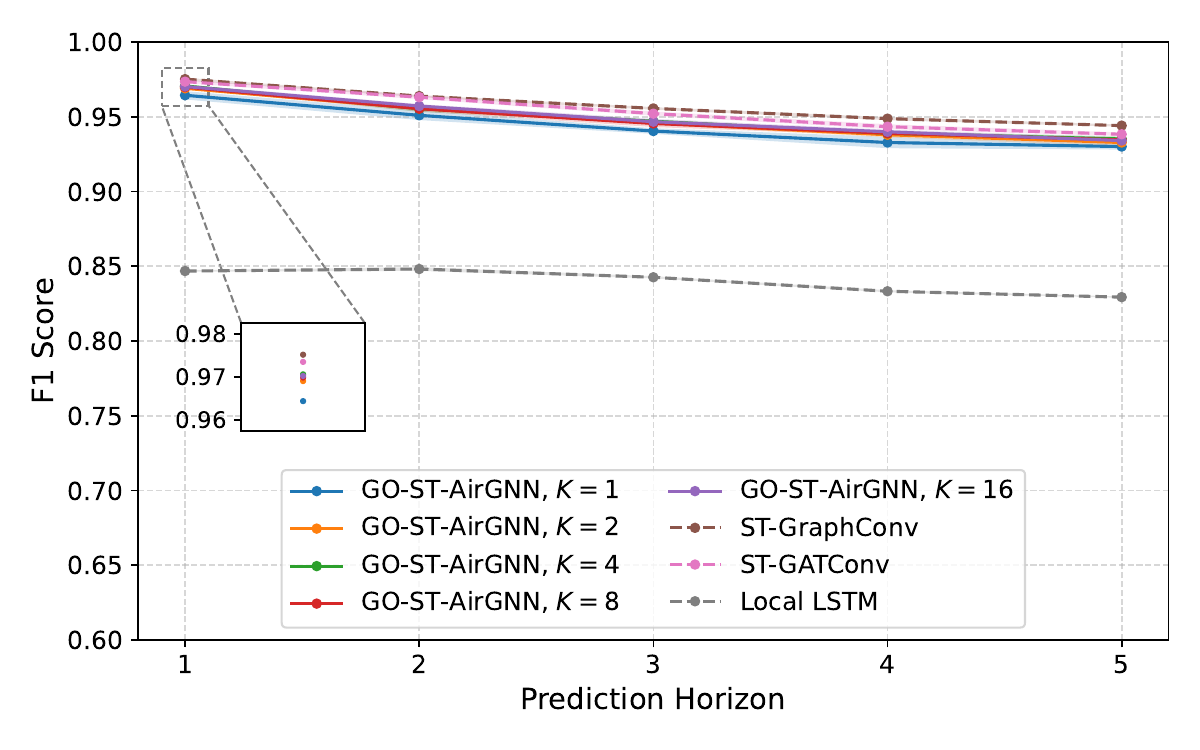}
        \caption{F1 score.}
    \end{subfigure}
    \caption{Performance evaluation of inductive generalization on unseen graph topologies.}
    \label{fig:scalability_results}
\end{figure*}

\subsubsection{Implementation and Hyperparameters}
The local transmission modules, i.e., the semantic encoder $\Phi_i$ and the resource manager $\Psi_i$, are realized via \acp{MLP}.
For the receiver-side semantic decoder $\chi_i$, we employ a $2$-layer \ac{LSTM} network with $64$ hidden units. This recurrent structure allows the node to effectively integrate the temporal history of the spatially aggregated signals.
The message embedding dimension is set to $d=64$.
To prioritize ultra-low latency inference, which is critical for the real-time \ac{IIoT} control loop, we restrict the architecture to a \textit{single} \ac{GNN} aggregation layer for both the proposed method and the baselines.
The model is trained end-to-end using the Adam optimizer. 
To ensure statistical reliability, all reported results represent the average of 10 independent trials with different random seeds.

\subsubsection{Baselines}
To validate the effectiveness of the proposed goal-oriented analog aggregation, we benchmark \ac{GO-ST-AirGNN} against three distinct architectures:

\begin{itemize}
    \item \textbf{Local \ac{LSTM}}. This baseline operates in a purely non-cooperative manner. Each node attempts to predict blockages relying solely on the temporal history of its own local features $\mathbf{x}_{i,t}$, without exchanging information with neighbors. This serves as a performance \textit{lower bound}, quantifying the fundamental gain yielded by spatial collaboration.

    \item \textbf{\Ac{ST-GCN}}. This baseline combines standard spectral graph convolution with the temporal LSTM decoder. It aggregates neighbor features treating all connected neighbors with equal importance. It represents a topology-agnostic digital approach.

    \item \textbf{\Ac{ST-GAT}}. This architecture employs digital attention mechanisms to assign importance weights to different neighbors. We include it to compare our proposed \textit{physical} weighting strategy against a standard \textit{digital} weighting approach, where neighbor relevance is computed computationally rather than via channel selection.
    
\end{itemize}

\textit{Remark on Digital Baselines}. It is crucial to interpret \ac{ST-GCN} and \ac{ST-GAT} as \textit{empirical reasoning upper bounds}, because they are granted perfect, instantaneous digital data exchange to assess the maximum possible prediction accuracy achievable with perfect information sharing. 
However, they ignore the physical reality of the wireless channel. In a real-world deployment, these digital methods would be subject to orthogonal access constraints, incurring in latency penalties, different from the proposed \ac{GO-ST-AirGNN}.

\subsection{Inductive Generalization}
\label{sec:scalability_results}

A key characteristic of graph-based learning is \textit{inductive generalization}, i.e., the ability of a model trained on specific graph structures in $\mathcal{T}_{\textbf{train}}$ to generalize to unseen topologies in $\mathcal{T}_{\textbf{test}}$. 
In the context of \ac{IIoT}, this property is crucial, as the deployment of active nodes in a facility often change dynamically.
In Fig. \ref{fig:scalability_results}, we present the quantitative results of our inductive generalization experiment in terms of accuracy and F1 score across prediction horizons ranging from 1 to 5 steps. 
A distinct performance gap is observed between the non-cooperative Local LSTM and all graph-based approaches. This confirms that blockage events in dynamic industrial environments cannot be reliably predicted solely from local history and spatial context from neighbors is essential to anticipate incoming blockages.

The proposed GO-ST-AirGNN demonstrates competitive performance, compared to the state-of-the-art graph baselines with ideal digital communication, while accounting for channel characteristics of actual physical environments. We remark that these state-of-the-art baselines overlook practical impairments of the wireless channel and assume sufficiently large communication resources [Section \ref{subsec:efficiency}]; hence, being considered as theoretical upper bounds only for reference. 
This indicates that our model is inherently able to deal with the superposition of analog signals for feature aggregation. 
Both accuracy and F1 scores of all approaches show a gradual decline as the prediction horizon $\tau$ increases from 1 to 5, corresponding to a more challenging prediction problem. GO-ST-AirGNN maintains satisfactory performance, and increasing the number of parallel frequencies $K$ yields improved results, which corroborate our theoretical findings that more communication resources provide stronger expressive power [Section \ref{sec:expressivity}].  
These results demonstrate that the proposed \ac{GO-ST-AirGNN} possesses a strong generalization capacity, while featuring significantly higher spectral efficiency.

\begin{figure}[t]
    \centering
    \includegraphics[width=0.9\linewidth]{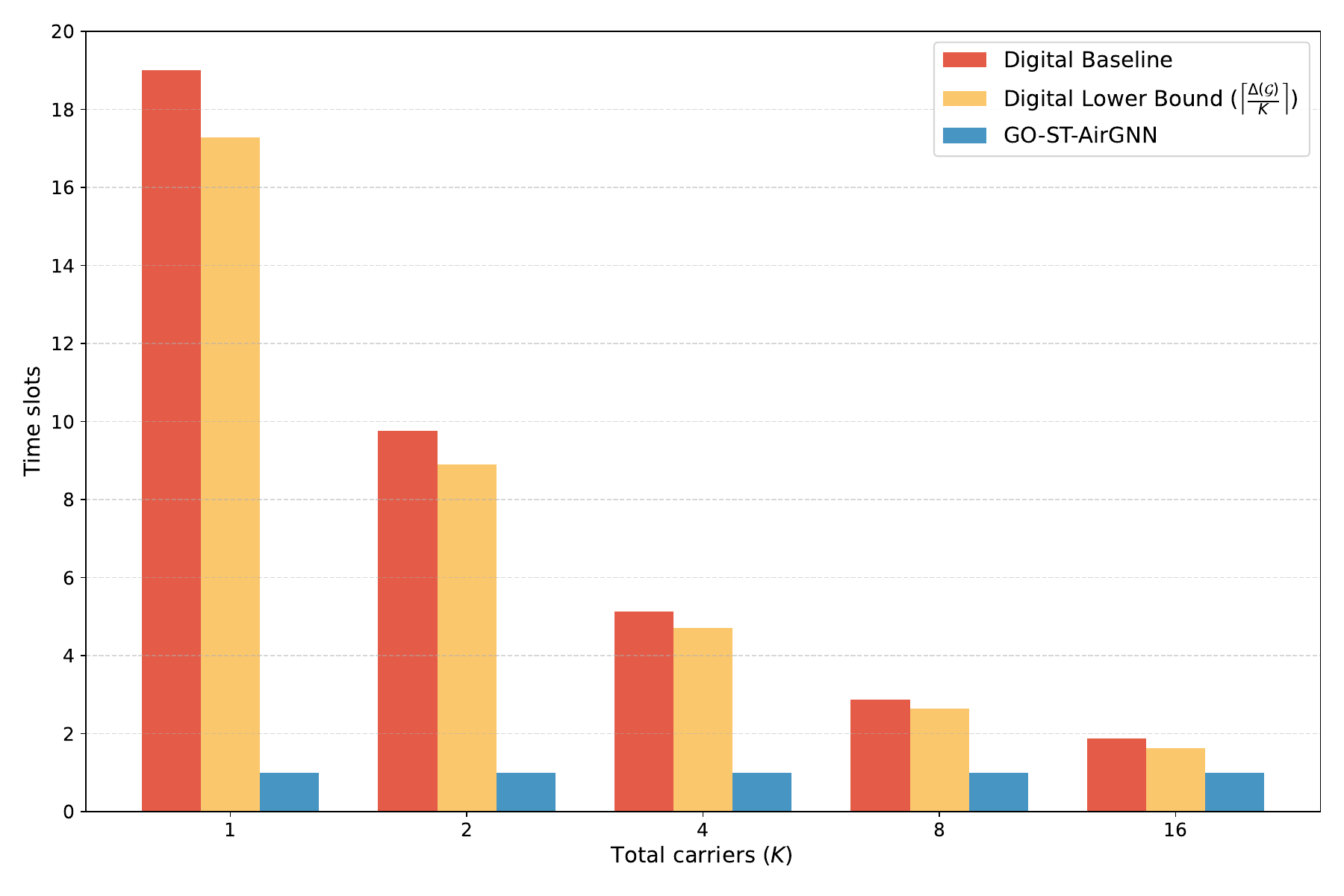} 
    \caption{Comparison of communication latency.}
    \label{fig:latency_comparison}
\end{figure}

\textit{Remark on Practical Expressivity and Latency.} It is important to emphasize that while the theoretical proof of expressivity (Proposition 2) relies on the availability of sufficient orthogonal resources to match representational capabilities of digital GNNs, empirical results shown in Fig.~\ref{fig:scalability_results} suggest that such strict orthogonality is not a prerequisite for effective task performance. GO-ST-AirGNN maintains high accuracy even when $K$ is small, exhibiting remarkable robustness to the number of subcarriers. %, maintaining high accuracy even when $K$ is small. 
This indicates that the architecture is inherently capable of processing non-orthogonal signal superpositions effectively. 
As shown in Fig.~\ref{fig:latency_comparison}, digital counterparts operating in this low-$K$ regime suffer from severe latency penalties. The figure illustrates this by plotting the actual digital performance alongside its theoretical lower bound. The latency of digital approaches scales linearly with the neighborhood size ($T_\text{dig} \propto \frac{\Delta(\mathcal{G})}{K}$). In contrast, GO-ST-AirGNN scales effectively, achieving constant use of a single time slot independent of network density.
Consequently, the proposed over-the-air aggregation proves to be more suitable for dense, resource-constrained networks than standard digital approaches.

\subsection{Robustness to Domain Shift}
\label{sec:online_learning}

\begin{figure}[t]
    \centering
    \includegraphics[width=0.9\linewidth]{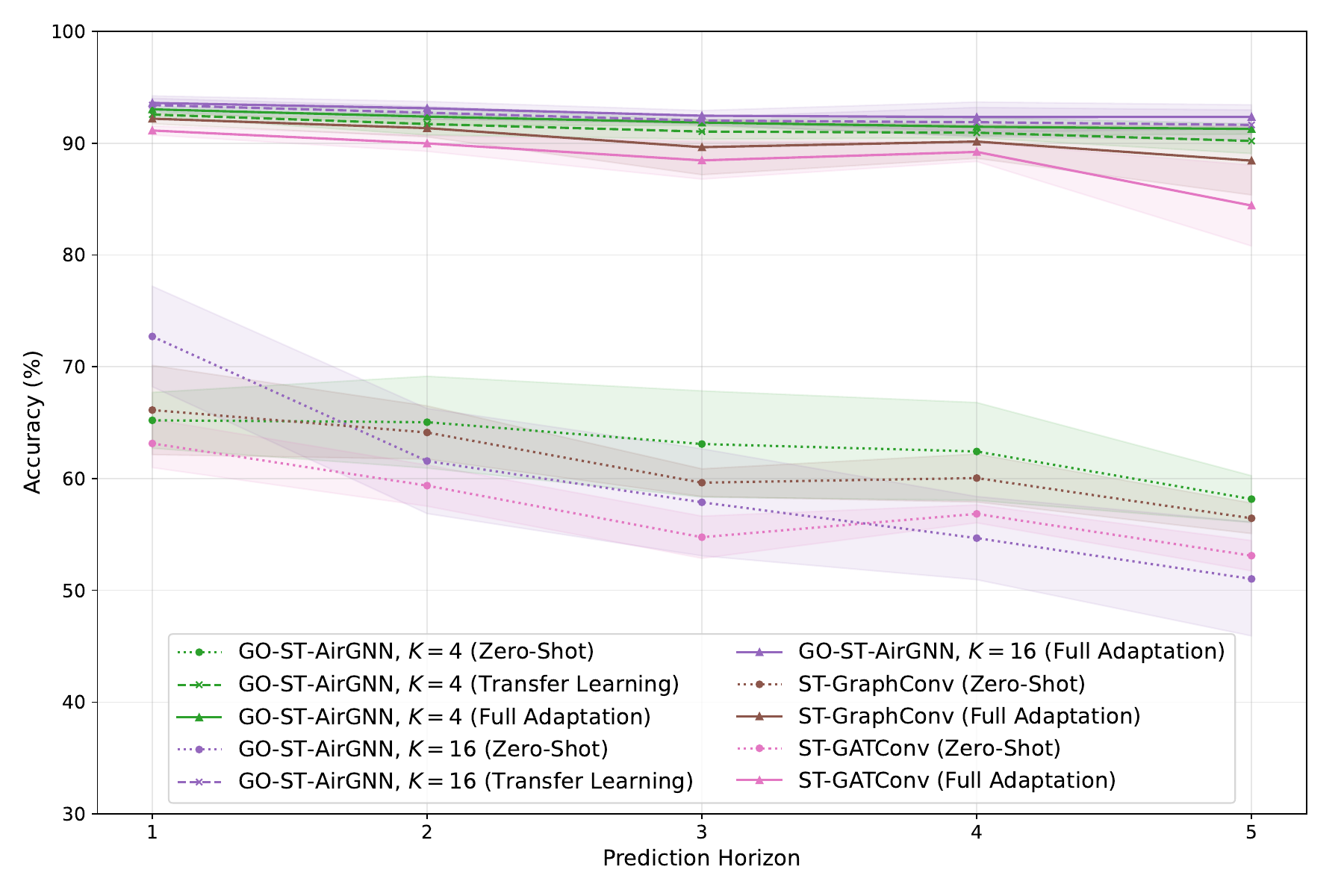} 
    \caption{Adaptation to domain shift.}
    \label{fig:online_learning}
\end{figure}

Next, we evaluate the robustness of \ac{GO-ST-AirGNN} to domain shifts, including different mobile blockers than those seen during training. We test the trained model on an additional dataset $\mathcal{T}'_{\text{test}}$, which features a mobile robot with a significantly different physical shape compared to the one in the training set. To simulate a realistic rapid deployment scenario, we assume access to a scarce adaptation dataset $\mathcal{T}'_{\text{adapt}}$ comprising only $10$ graph time series, in sharp contrast to the $1000$ sequences available in $\mathcal{T}_{\text{train}}$ and $\mathcal{T}'_{\text{test}}$.
For clarity, we present the performance of \ac{GO-ST-AirGNN} with $K=\{4, 16\}$ and digital graph baselines. 
Fig.~\ref{fig:online_learning} reports the accuracy across varying prediction horizons under three operational modes:

\begin{enumerate}
    \item \textbf{Zero-Shot}. The model trained on the original environment is applied directly to the new domain without any updates.
    \item \textbf{Transfer Learning}. The semantic encoder modules ($\Phi$ and $\Psi$) are \textit{frozen}. Only the receiver-side semantic decoder $\chi$ is fine-tuned on the new data. This represents a lightweight adaptation strategy where the communication protocol remains fixed.
    \item \textbf{Full Adaptation}. The entire end-to-end pipeline ($\Phi, \Psi, \chi$) is fine-tuned on the new domain.
\end{enumerate}

\begin{figure*}[t]
    \centering
    % First Subfigure: Strategy Impact
    \begin{subfigure}[b]{0.34\linewidth}
        \includegraphics[width=\linewidth]{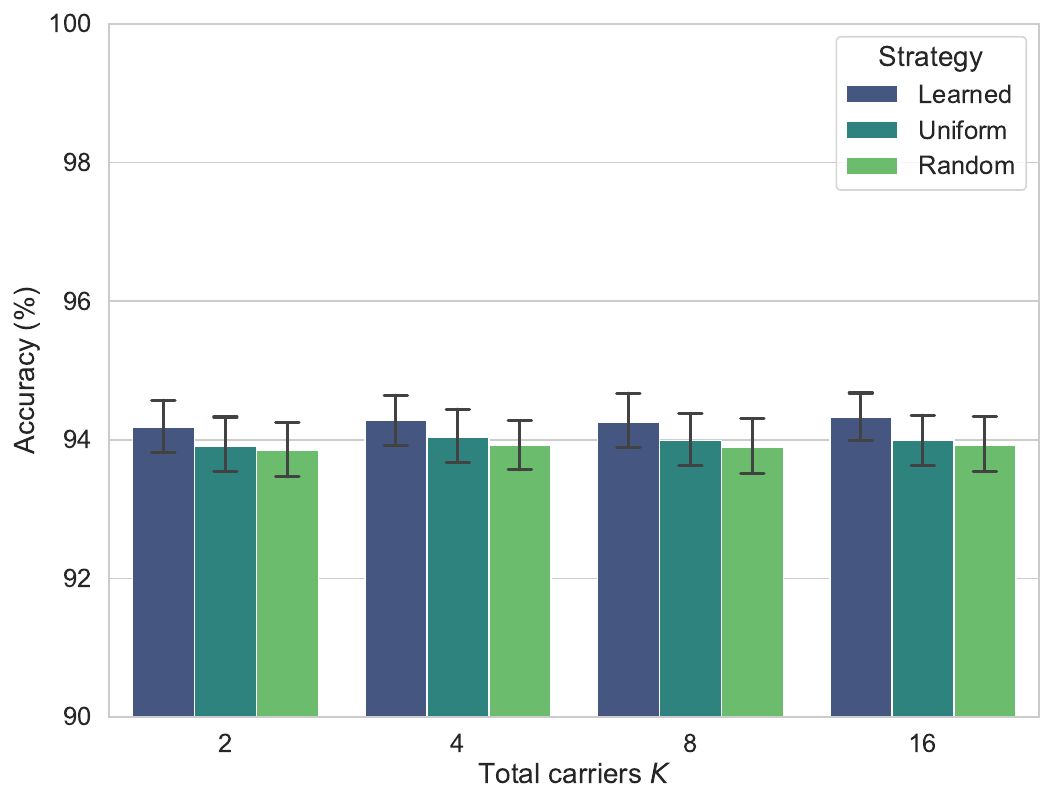}
        \caption{Impact of allocation strategies}
        \label{fig:strategy_impact}
    \end{subfigure}
    \quad \quad
    % Second Subfigure: Budget Sensitivity
    \begin{subfigure}[b]{0.34\linewidth}
        \includegraphics[width=\linewidth]{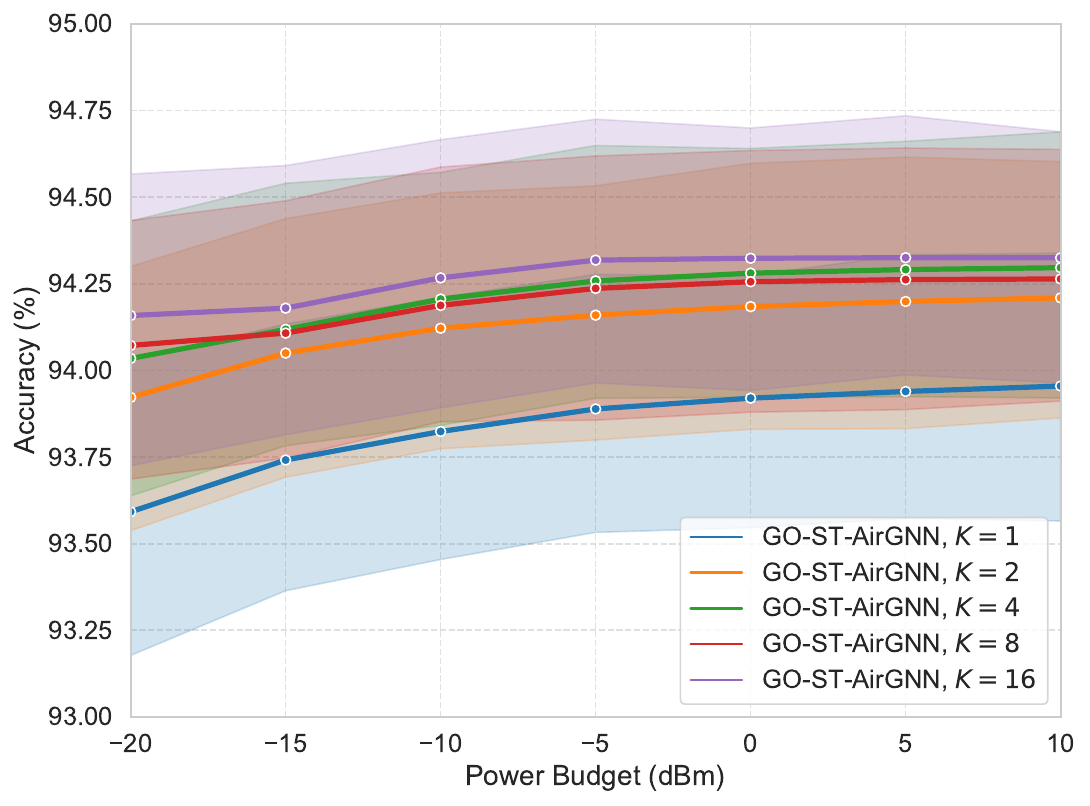}
        \caption{Sensitivity to power budget}
        \label{fig:budget_sensitivity}
    \end{subfigure}
    
    \caption{Ablation study on goal-oriented power management.}
    \label{fig:power_strategies}
\end{figure*}

\begin{figure*}[t]
    \centering
    % First Subfigure: Short Horizon
    \begin{subfigure}[b]{0.34\linewidth}
        \includegraphics[width=\linewidth]{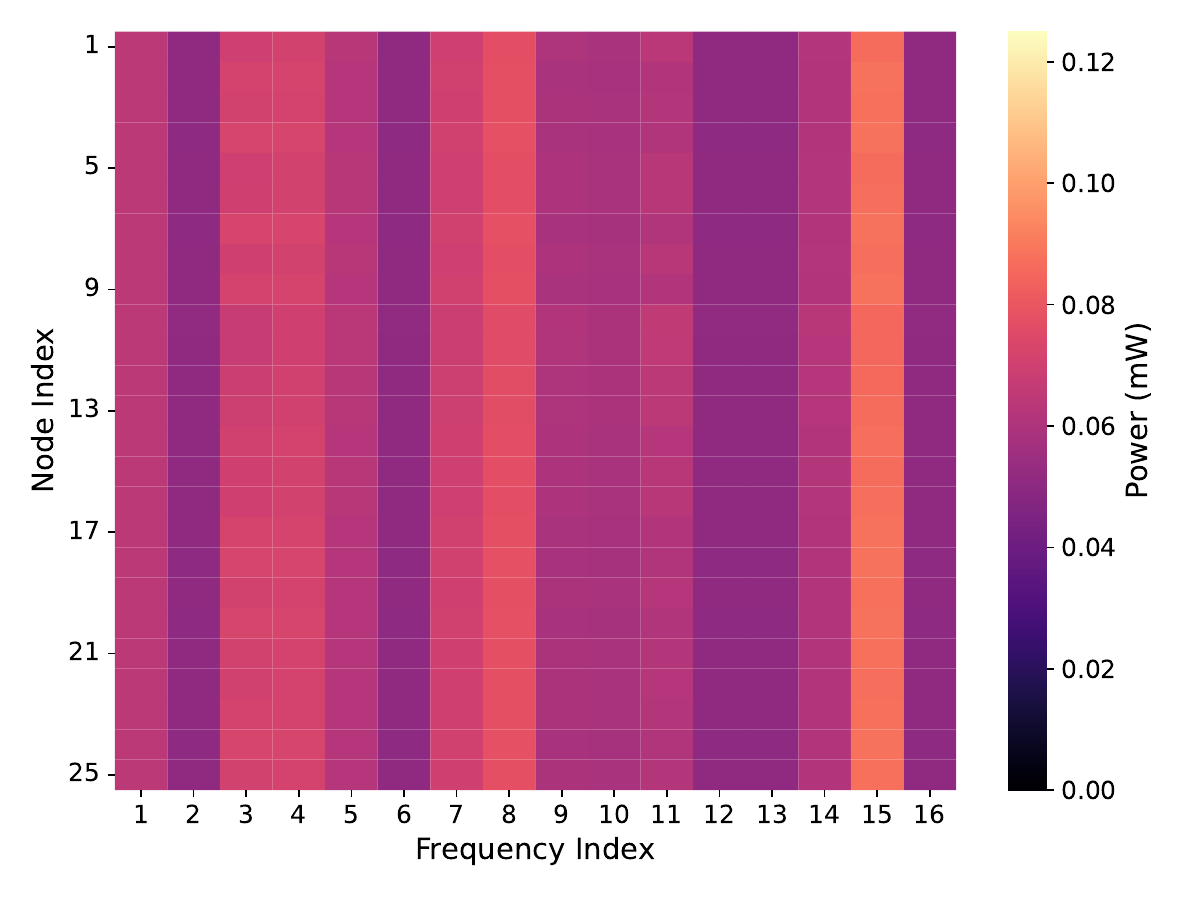}
        \caption{Short horizon ($\tau = 1$)}
        \label{fig:heatmap_h1}
    \end{subfigure}
    \quad \quad 
    % Second Subfigure: Long Horizon
    \begin{subfigure}[b]{0.34\linewidth}
        \includegraphics[width=\linewidth]{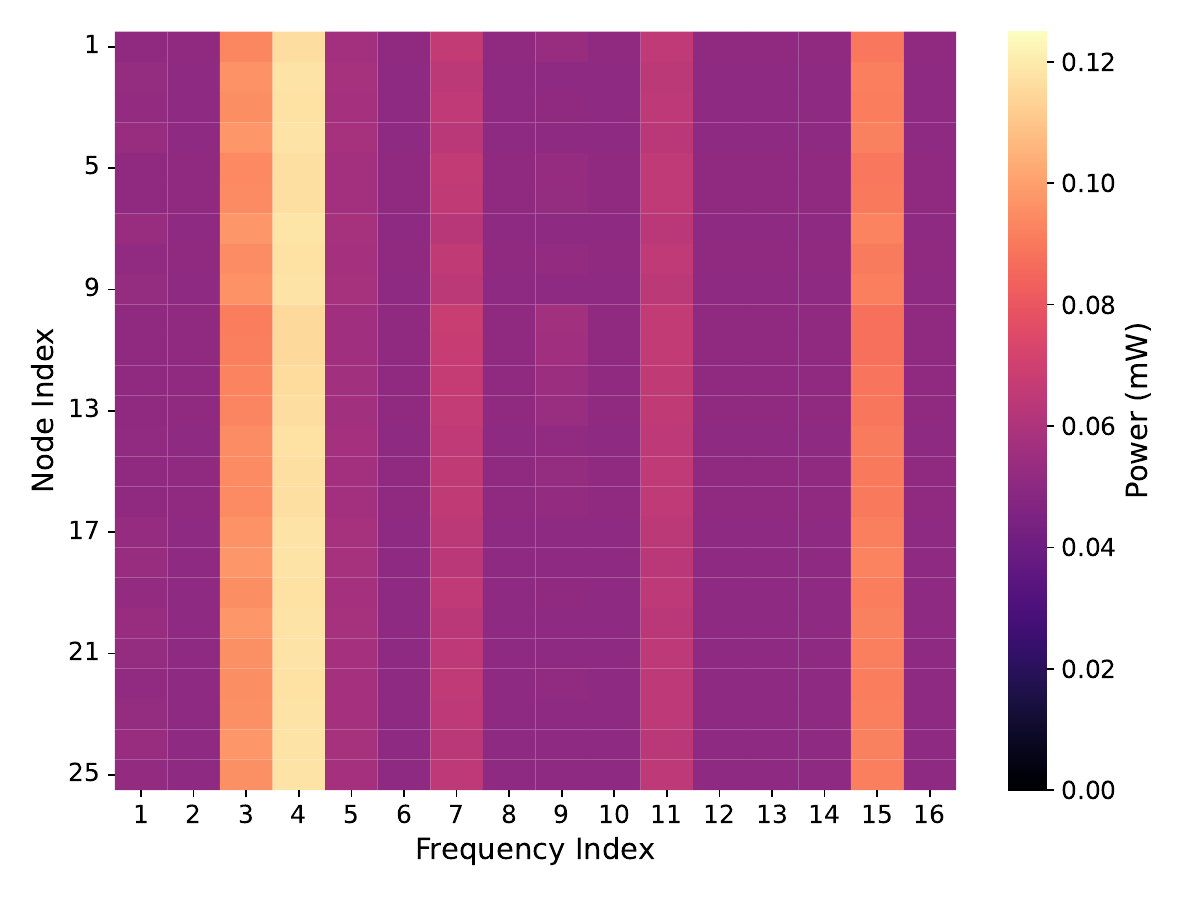}
        \caption{Long horizon ($\tau = 5$)}
        \label{fig:heatmap_h5}
    \end{subfigure}
    
    \caption{
    Visualization of learned goal-oriented power allocation strategies, suggesting task-adaptive topology shaping.}
    \label{fig:power_heatmaps}
\end{figure*}

\noindent First, we observe that the \textit{Zero-Shot} performance naturally degrades due to the mismatch in blockage signatures between the training and test environments. 
However, 
the \textit{Transfer Learning} strategy of the proposed \ac{GO-ST-AirGNN} proves exceptionally effective. 
By fine-tuning only the local semantic decoder $\chi_i$ on the scarce adaptation dataset $\mathcal{T}'_{\text{adapt}}$ (while keeping the transmission policy $\Phi_i, \Psi_i$ frozen), the system recovers nearly the entire performance gap, i.e., the accuracy of the transfer learning approach converges to that of the \textit{Full Adaptation}, where the entire pipeline is retrained. 
This implies a fundamental decoupling in the learned architecture: the \textit{encoder} learns a robust and general communication strategy that remains valid across different environments, while the \textit{decoder} handles the specific interpretation of the local blockage dynamics. It allows for rapid, low-complexity adaptation at the edge without requiring network-wide re-calibration. 
Second, as illustrated in Fig.~\ref{fig:online_learning}, \ac{GO-ST-AirGNN} with either transfer learning or full adaption outperforms \ac{ST-GCN} and \ac{ST-GAT} with full adaption, demonstrating that the proposed framework delivers stronger robustness and adaptability compared to digital graph baselines. 
These results confirm that our goal-oriented analog approach achieves competitive or even better performance than digital baselines under domain shifts.
Moreover, these results further confirm that \ac{GO-ST-AirGNN} provides an effective strategy to reduce communication cost while exhibiting competitive performance.

\subsection{Analysis of Goal-Oriented Power Allocation}

In this section, we interpret how \ac{GO-ST-AirGNN} utilizes the spectral domain according to its goal-oriented resource management policy. To isolate the specific contribution of the goal-oriented resource management module $\Psi(\cdot)$, we conduct an ablation study in Fig. \ref{fig:power_strategies}. For conciseness, the results depicted in this figure are averaged over all considered prediction horizons $\tau \in \{1, \dots, 5\}$.
We compare the \textit{Learned} power allocation policy against two heuristic baselines:
(i) \textit{Uniform}, where the power budget $P_{\text{tot}}$ is distributed equally across all $K$ subcarriers; and
(ii) \textit{Random}, where the power allocation is sampled from a uniform distribution subject to the budget constraint.
As illustrated in Fig. \ref{fig:power_strategies}(a), the \textit{Learned} strategy consistently yields the highest prediction accuracy across all configurations of $K$, confirming that learning the underlying topologies of communication graphs %topologies 
aids the aggregation process. 
In Fig. \ref{fig:power_strategies}(b), we analyze the sensitivity of the system to the maximum transmission power budget $P_{\text{tot}}$, varying the constraint from $-20$ dBm to $10$ dBm. The accuracy curves exhibit a steeper improvement in the low-power regime followed by a plateau starting at approximately $-5$ dBm, indicating that the system can operate in an energy-efficient regime without compromising prediction reliability.

\begin{figure*}[t]
    \centering
    % --- Row 1 ---
    \begin{subfigure}[b]{0.32\textwidth}
        \includegraphics[width=\linewidth]{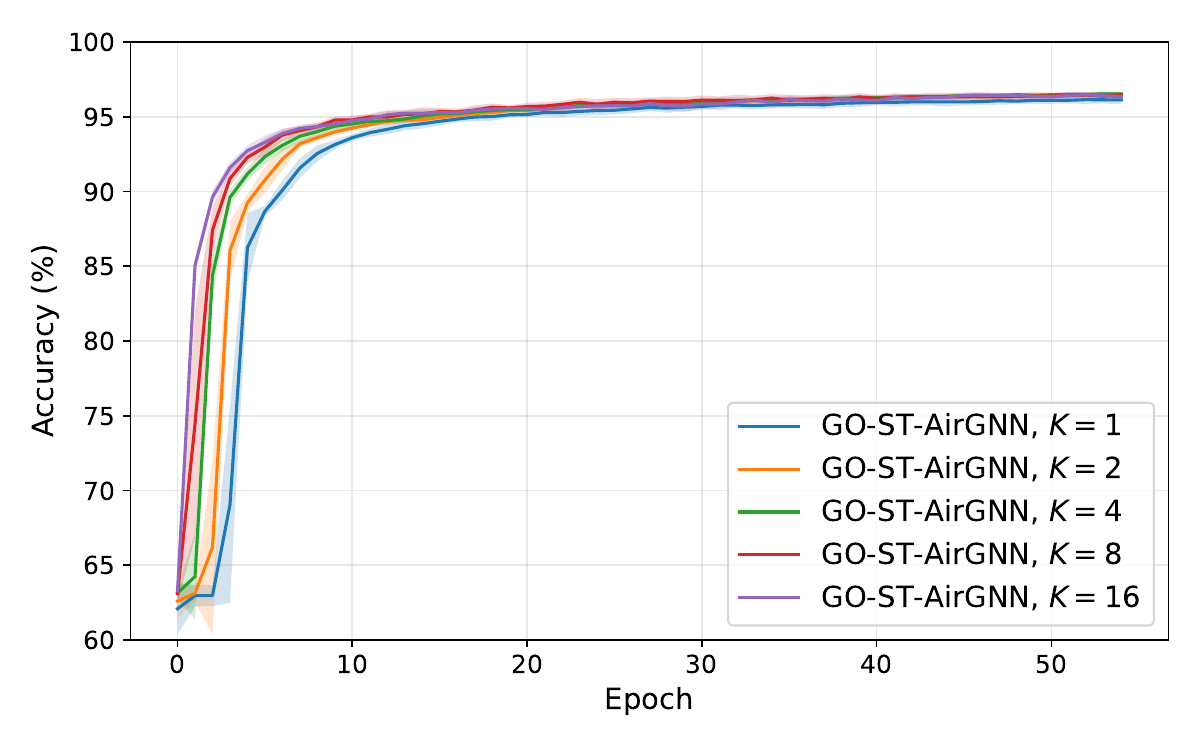}
        \caption{Accuracy, $\tau=1$}
    \end{subfigure}
    \hfill
    \begin{subfigure}[b]{0.32\textwidth}
        \includegraphics[width=\linewidth]{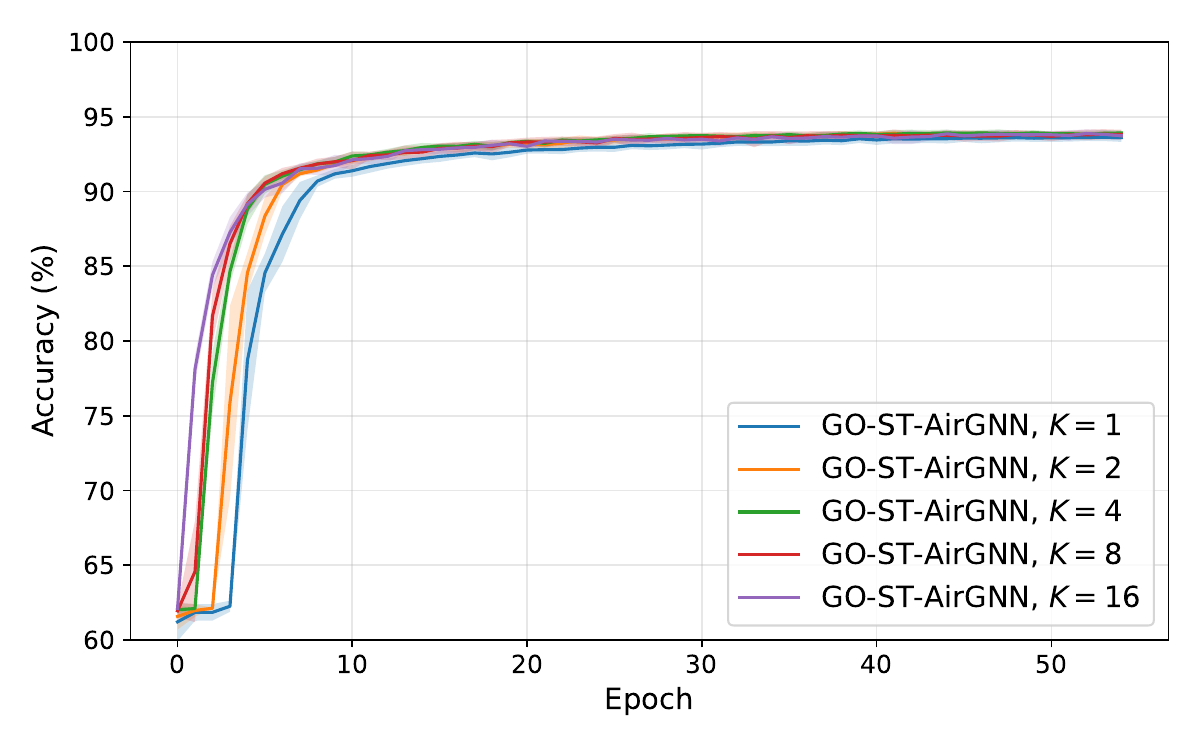}
        \caption{Accuracy, $\tau=3$}
    \end{subfigure}
    \hfill
    \begin{subfigure}[b]{0.32\textwidth}
        \includegraphics[width=\linewidth]{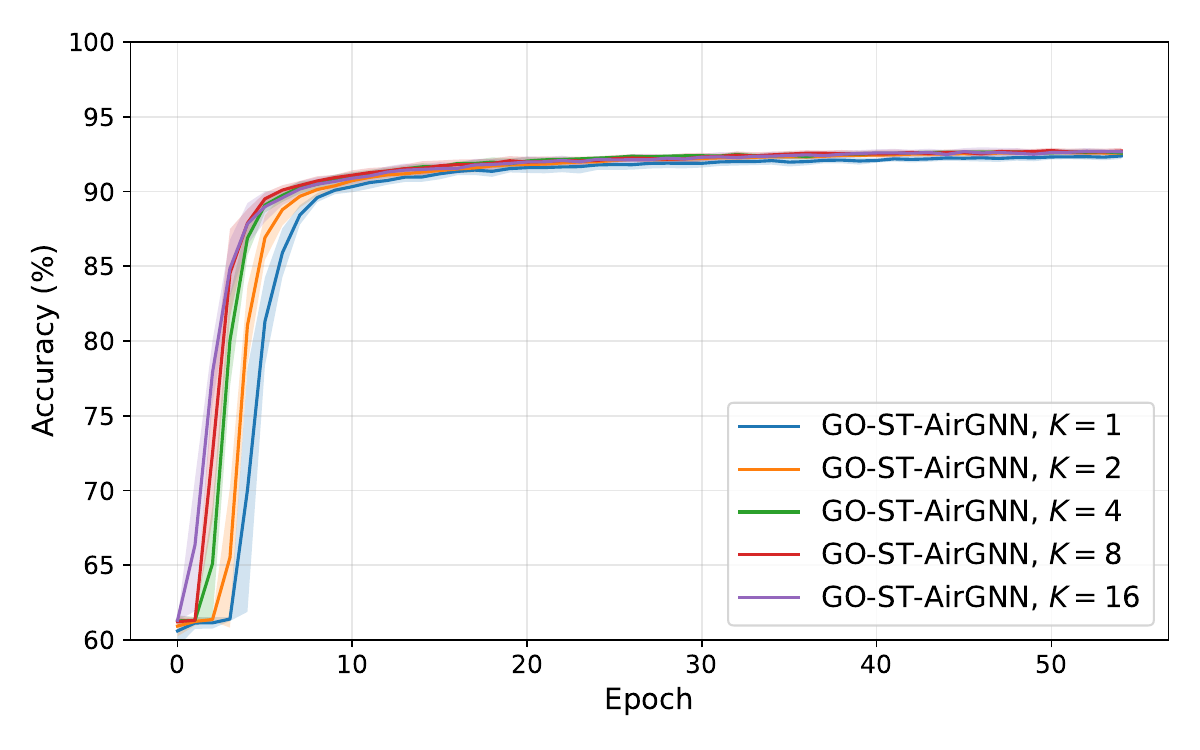}
        \caption{Accuracy, $\tau=5$}
    \end{subfigure}

   \caption{Training convergence analysis.}
   \label{fig:training_convergence}
\end{figure*}

To understand how the goal-oriented policy $\Psi(\cdot)$ optimizes the graph topology, we visualize the power allocation matrices in Fig. \ref{fig:power_heatmaps}. Considering GO-ST-AirGNN operating with $K=16$, the heatmaps depict examples of transmitting power levels allocated by $N=25$ nodes across the available subcarriers for $\tau=1$ and $\tau=5$.
A common feature across both scenarios is that the allocation is generally non-sparse; nodes tend to utilize the entire spectral budget rather than selecting a single subcarrier, suggesting that \ac{GO-ST-AirGNN} leverages the diversity of the $K$ parallel graphs to build a robust ensemble, reinforcing local connectivity.
Comparing Fig. \ref{fig:heatmap_h1} ($\tau=1$) and Fig. \ref{fig:heatmap_h5} ($\tau=5$) reveals a clear shift in strategy based on the difficulty of the prediction task.
For the former case, the power distribution is relatively balanced. Since the blockage event is imminent, strong local correlations are sufficient. In the latter case, instead, as the prediction horizon extends, the causal relationship between current features and future blockages becomes more complex and often requires capturing spatially distant dynamics (e.g., an approaching robot detected by a far neighbor). To address this, the policy shifts to allocate more power to a specific subcarrier, visible as bright vertical yellow bands in Fig. \ref{fig:heatmap_h5}, facilitating reliable longer-range information exchanges that are critical for anticipating future system states.
This behavior empirically demonstrates the goal-oriented nature of the architecture: the physical graph topology is not static, but is dynamically reshaped to match the specific spatio-temporal requirements of the downstream prediction task. Therefore, the communication strategy (i.e., message passing operations) is learned contingent on the learning goal.

\subsection{Training Convergence Analysis}

While the considered CTDE paradigm relies on offline training, evaluating the convergence trends of \ac{GO-ST-AirGNN} provides critical insights into its learning efficiency. Fig. \ref{fig:training_convergence} shows training convergence trends. Notably, increasing the number of subcarriers $K$ from 2 to 16 significantly enhances convergence speed across all considered scenarios ($\tau={1,3,5}$). This acceleration can be attributed to the increased spectral diversity and resource granularity provided by a larger $K$. These additional degrees of freedom effectively smooth the optimization landscape, allowing the system to identify optimal topological configurations more rapidly. Overall, all variants exhibit highly stable training behavior, consistently reaching convergence plateaus within just 15-20 epochs, validating the method's robustness.

\section{Conclusion}
\label{sec:conclusions}

In this paper, we introduced \ac{GO-ST-AirGNN}, a goal-oriented communication framework that integrates \ac{AirComp} with spatio-temporal graph learning to address the critical challenge of \ac{LOS} blockage prediction in dense \ac{IIoT} environments. 
Unlike digital methods, \ac{GO-ST-AirGNN} utilizes the wireless channel as an analog processor and exploits wireless signal superposition to conduct feature aggregation %operate 
effectively in non-orthogonal regimes. 
% In the specific context of LOS blockage prediction, wireless impairments may actually carry semantic meaning. Consequently, GO-ST-AirGNN does not need to compensate for all channel effects, but only for those that are semantically uninformative regarding the downstream task.
%
Our theoretical and empirical analysis yielded three key insights:
\begin{itemize}
    \item The proposed architecture offers a decisive scalability advantage, achieving constant communication cost independent of node density, whereas digital baselines scale linearly with the neighborhood size of the communication graph.
    \item The expressivity of the analog aggregation via AirComp is comparable to its digital counterparts, providing a universal approximation in terms of expressive capacity.
    \item High-fidelity simulations demonstrated that the learned policies of GO-ST-AirGNN exhibit strong inductive generalization to unseen network topologies 
    and can adapt to domain shifts via lightweight transfer learning.
\end{itemize}
It is worth noting that %Crucially, we observed that 
GO-ST-AirGNN achieves competitive performance even with very limited number of carriers, effectively unlocking scalability gains.

Looking forward, these findings open a promising avenue for establishing low-latency and spectral-efficient networked systems, where AirComp leverages superposition of analog signals for feature aggregation and goal-oriented communications align resource management with decision-making strategies for %of %the intrinsic complexity of 
%the 
downstream tasks.
Additionally, future research will explore the application of our framework to distinct sensing tasks that impose unique constraints on graph topology shaping, alongside an extension to distributed training schemes.

\appendix

\label{appendix:A}

\textit{Proof of Proposition 1.}
The proof relies on the analysis of radio resource constraints in both schemes.

\textit{1) \ac{GO-ST-AirGNN}:}
In the proposed scheme, the aggregation target for node $i$ over the $k$-th resource is the weighted sum $\tilde{\mathbf y}_{i,k} = \sum_{j \in \mathcal{D}_i} \mathbf h_{j,i,k}\ \tilde{\mathbf{x}}_{j,k}.$ This operation is mapped directly to the superposition property of the wireless channel, where every node $j \in \mathcal{N}$ transmits its feature vector $\mathbf{x}_j$ simultaneously over resource $k$.
Since the channel naturally performs the summation over-the-air, interference is treated as the aggregation operator rather than a collision. Consequently, all neighborhood aggregations for all $i \in \mathcal{N}$ are completed in a single time slot. It follows that $T_{\text{air}} = 1.$

\textit{2) Digital \ac{GNN}:}
In a digital scheme, node $i$ must decode individual messages from every neighbor $u \in \mathcal{D}_i$ to compute the aggregation digitally. This imposes a strict \textit{orthogonality constraint}: signals from distinct neighbors $u,v \in \mathcal{D}_i$ targeting the same receiver $i$ must be orthogonal in time or frequency to avoid decoding failure.
We formalize this resource allocation as a \textit{graph coloring problem} \cite{pardalos1998graph}. Let $\mathcal{G}_c = (\mathcal{N}, \mathcal{E}_c)$ be the \textit{conflict graph} of the network. 
Two nodes $u$ and $v$ share an edge in the conflict graph $\mathcal{E}_c$ if and only if they are distinct and share a common receiver (i.e., they are both neighbors of the same node $i$), i.e.,
\begin{equation}
\mathcal{E}_c = \bigcup_{i \in \mathcal{N}} \left\{ (u, v) \mid u, v \in \mathcal{D}_i, u \neq v \right\},
\end{equation}
where $\bigcup(\cdot)$ is the union operator. 
The orthogonality constraint implies that for any receiver $i$, all its neighbors $\mathcal{D}_i$ must be assigned distinct resources.
Mathematically, the subgraph induced by $\mathcal{D}_i$ in $\mathcal{G}_c$ forms a \textit{clique} of size $|\mathcal{D}_i|$, i.e., a graph where every node is connected to every other node.
A fundamental property of graph coloring is that the minimum number of resources $\rho(\mathcal{G}_c)$ needed to achieve orthogonal resource assignments for $\mathcal{D}_i$ is lower-bounded by the size of the largest clique $\omega(\mathcal{G}_c)$ in the graph, because all nodes in a clique are mutually adjacent and must therefore be assigned distinct resources. It follows that 
\begin{equation}
    \rho(\mathcal{G}_c) \ge \omega(\mathcal{G}_c) = \max_{i} |\mathcal{D}_i| = \Delta(\mathcal{G}).
\end{equation}
Given a bandwidth budget of $K$ parallel subcarriers per time slot, the number of time slots required is:
\begin{equation}
    T_{\text{dig}} = \left\lceil \frac{\rho(\mathcal{G}_c)}{K} \right\rceil \ge \left\lceil \frac{\Delta(\mathcal{G})}{K} \right\rceil.\quad\square
\end{equation}

\vspace{0.2em}
\textit{Proof of Proposition 2.}

The proof relies on the capacity of \ac{GO-ST-AirGNN} to orthogonalize communication links in the frequency domain at any time slot $t$. While $K$ parallel channel realizations exist between any node pair $(i,j)$, we need only \textit{one} active subcarrier to convey the message $\mathcal{M}(\mathbf{u}_i, \mathbf{u}_j)$. If $K$ is large enough, we can assign distinct frequencies to distinct edges, preventing the channel from forcing an over-the-air aggregation and allowing the decoder to recover individual messages.

\textit{1) Frequency Assignment:}
As discussed in the proof of Proposition 1, this resource allocation problem can be framed as a graph coloring problem. If $\mathcal{G}_c$ is conflict graph of the network and $\rho(\mathcal{G}_c)$ is the minimum number of resources needed to satisfy the orthogonality constraint, it follows that $K' = \rho(\mathcal{G}_c)$ is the number of subcarriers needed by a standard \ac{MPNN} update rule to assign orthogonal frequencies to active links.
Since $K = K'$, there exists an injective mapping $\nu: \mathcal{E} \to \{1, \dots, K\}$ that assigns an interference-free subcarrier $k^* = \nu(j,i)$ to every directed link $(j,i)$. This ensures that no two active transmissions in the network occupy the same frequency.

\textit{2) Transmission Strategy:}
Given phase synchronization, the effective channel is represented by its magnitude $|\mathbf h_{j,i,k^*}|$. To achieve deterministic message transfer, the system must invert these channel gains.
First, the semantic encoder $\Phi_j$ constructs a unit-power message embedding $\mathbf{m}_{j}$ carrying the semantic content:
\begin{equation}
    [\mathbf{m}_{j}]_{k^*} = \mathcal{M}(\mathbf{u}_i, \mathbf{u}_j), \quad \text{with } |[\mathbf{m}_{j}]_{k^*}|^2 = 1.
\end{equation}
Second, the resource management network $\Psi_j$ allocates transmit power $p_{j, k^*}$ to invert the channel gain:
\begin{equation}
    p_{j, k^*} = \frac{P_{\text{rx}}}{|\mathbf h_{j,i,k^*}|},
\end{equation}
where $P_{\text{rx}}$ is a target received signal power. This inversion is feasible provided the required power is within the budget, i.e., $p_{j, k^*} \leq P_{\text{tot}}$.

\textit{3) Analog Reception with Noise:}
At receiver $i$, the signal on the unique assigned subcarrier $k^*$ is the product of the channel gain, the transmit power scaling, and the message embedding. Due to the injectivity of $\nu$, there is no interference. The received signal is:
\begin{align}
    \tilde{\mathbf y}_{i, k^*} &= |\mathbf  h_{j,i,k^*}| \cdot {p_{j, k^*}} \cdot [\mathbf{m}_{j}]_{k^*} + \mathbf z_{i,k^*} \nonumber \\
    &= |\mathbf h_{j,i,k^*}| \cdot \frac{P_{\text{rx}}}{|\mathbf h_{j,i,k^*}|} \cdot \mathcal{M}(\mathbf{u}_i, \mathbf{u}_j) + \mathbf z_{i,k^*} \nonumber \\
    &= P_{\text{rx}} \cdot \mathcal{M}(\mathbf{u}_i, \mathbf{u}_j) + \mathbf z_{i,k^*},
    \label{eq:noisy_approx}
\end{align}
where $\mathbf z_{i,k^*} \sim \mathcal{CN}(0, \sigma^2)$ is the thermal noise.

\textit{4) Semantic Decoding:}
The received vector $\mathbf{y}_i$ now contains the separated messages $\{\mathcal{M}(\mathbf{u}_i, \mathbf{u}_j) \}_{j \in \mathcal{N}_i}$ scaled by $P_{\text{rx}}$, plus noise. The semantic decoder $\chi_i$, acting as a universal function approximator, can be trained to:
(a) Normalize the input entries by $1/P_{\text{rx}}$ to recover the original message scale (while ignoring inactive subcarriers).
(b) Apply the desired aggregation operator $\bigoplus$ (e.g., max, mean) over the recovered messages.
(c) Apply the update function $\mathcal{U}$.

Eq. \ref{eq:noisy_approx} reveals that the over-the-air reconstruction is inherently stochastic due to thermal noise $\mathbf z_{i,k^*}$. However, the approximation becomes exact in the high-\ac{SNR} regime. Specifically, if the power budget $P_{\text{tot}}$ is sufficient such that $\frac{P_{\text{tot}}}{\sigma^2} \to \infty$, the effect of $\mathbf z_{i,k^*}$ becomes negligible relative to the signal, and \ac{GO-ST-AirGNN} converges to the deterministic MPNN behavior. 
Thus, the output $\hat{y}_{i, t+\tau}$ approximates the generic \ac{MPNN} layer, proving the proposition.
$\square$

\begingroup
\small
\bibliographystyle{IEEEtran}
\bibliography{references}

@STRING{Globecom = {Proc. IEEE Global Telecomm. Conf.}}

@STRING{ICASSP = {Proc. IEEE Int. Conf. Acoustics, Speech, and Signal Processing (ICASSP)}}

@STRING{ICC = {Proc. IEEE Int. Conf. on Commun.}}

@STRING{IEEE = {The Institute of Electrical and Electronics Engineers,}}

@STRING{WCNC = {Proc. IEEE {W}ireless {C}ommun. and {N}etworking {C}onf. (WCNC)}}

@STRING{AI = {Artificial Intelligence (AI)}}

@STRING{JIN = {J. of the Institute of Navigation}}

@INPROCEEDINGS{SkocajBlocakge,
  author={Skocaj, Marco and Zugno, Tommaso and Blumenstein, Jiri and Boban, Mate},
  booktitle={2024 IEEE Globecom Workshops (GC Wkshps)}, 
  title={Line of Sight Blockage Prediction via Spatio-Temporal Graph Neural Networks}, 
  year={2024},
  volume={},
  number={},
  pages={1-6},
  doi={10.1109/GCWkshp64532.2024.11100946}}

@ARTICLE{GaoAirgnns,
  author={Gao, Zhan and Gündüz, Deniz},
  journal={IEEE Transactions on Signal Processing}, 
  title={Graph Neural Networks Over the Air for Decentralized Tasks in Wireless Networks}, 
  year={2025},
  volume={73},
  number={},
  pages={721-737},
  doi={10.1109/TSP.2025.3534685}}

@article{zhang2025intelligent,
  title={Intelligent integrated sensing and communication: a survey},
  author={Zhang, Jifa and Lu, Weidang and Xing, Chengwen and Zhao, Nan and Al-Dhahir, Naofal and Karagiannidis, George K and Yang, Xiaoniu},
  journal={Science China Information Sciences},
  volume={68},
  number={3},
  pages={131301},
  year={2025},
  publisher={Springer}
}

@inproceedings{gao2023decentralized,
  title={Decentralized channel management in {WLANs} with graph neural networks},
  author={Gao, Zhan and Shao, Yulin and G{\"u}nd{\"u}z, Deniz and Prorok, Amanda},
  booktitle={ICC 2024 - IEEE International Conference on Communications},
  pages={3072--3077},
  year={2023}
}

@article{gao2022wide,
  title={Wide and deep graph neural network with distributed online learning},
  author={Gao, Zhan and Gama, Fernando and Ribeiro, Alejandro},
  journal={IEEE Transactions on Signal Processing},
  volume={70},
  pages={3862--3877},
  year={2022},
  publisher={IEEE}
}

@ARTICLE{Gunduz2021Effective,
  author={Tung, Tze-Yang and Kobus, Szymon and Roig, Joan Pujol and Gündüz, Deniz},
  journal={IEEE Journal on Selected Areas in Communications}, 
  title={Effective Communications: A Joint Learning and Communication Framework for Multi-Agent Reinforcement Learning Over Noisy Channels}, 
  year={2021},
  volume={39},
  number={8},
  pages={2590-2603},
  doi={10.1109/JSAC.2021.3087248}}

@article{saad2019vision,
  title={A vision of {6G} wireless systems: Applications, trends, technologies, and open research problems},
  author={Saad, Walid and Bennis, Mehdi and Chen, Mingzhe},
  journal={IEEE network},
  volume={34},
  number={3},
  pages={134--142},
  year={2019},
  publisher={IEEE}
}

@article{shastri2022review,
  title={A review of millimeter wave device-based localization and device-free sensing technologies and applications},
  author={Shastri, Anish and Valecha, Neharika and Bashirov, Enver and Tataria, Harsh and Lentmaier, Michael and Tufvesson, Fredrik and Rossi, Michele and Casari, Paolo},
  journal={IEEE Communications Surveys \& Tutorials},
  volume={24},
  number={3},
  pages={1708--1749},
  year={2022},
  publisher={IEEE}
}

@article{zhou20246,
  title={{6-DoF} Location-and-Pose Estimation Toward Integrated Visible Light Communication and Sensing: Algorithm Design and Performance Limits},
  author={Zhou, Bingpeng and Wang, Xin and Shen, Yuan and Fan, Pingzhi},
  journal={IEEE Transactions on Signal Processing},
  volume={72},
  pages={2576--2593},
  year={2024},
  publisher={IEEE}
}

@article{gao2019parallel,
  title={Parallel end-to-end autonomous mining: An {IoT}-oriented approach},
  author={Gao, Yu and Ai, Yunfeng and Tian, Bin and Chen, Long and Wang, Jian and Cao, Dongpu and Wang, Fei-Yue},
  journal={IEEE Internet of Things Journal},
  volume={7},
  number={2},
  pages={1011--1023},
  year={2019},
  publisher={IEEE}
}

@article{das2023ambit,
  title={Ambit-process-based spatial-wideband {MIMO} channel model for sub-{THz} urban microcellular communication},
  author={Das, Shrayan and Sen, Debarati and Viterbo, Emanuele and Chavva, Ashok Kumar Reddy and Sharma, Diwakar and Nigam, Anshuman},
  journal={IEEE Transactions on Wireless Communications},
  volume={23},
  number={1},
  pages={559--574},
  year={2023},
  publisher={IEEE}
}

@article{chen2023channel,
  title={Channel measurement, characterization, and modeling for terahertz indoor communications above 200 {GHz}},
  author={Chen, Yi and Han, Chong and Yu, Ziming and Wang, Guangjian},
  journal={IEEE Transactions on Wireless Communications},
  volume={23},
  number={6},
  pages={6518--6532},
  year={2023},
  publisher={IEEE}
}

@article{yang2024blockage,
  title={Blockage-aware robust beamforming in {RIS}-aided mobile millimeter wave {MIMO} systems},
  author={Yang, Yan and Dang, Shuping and Wen, Miaowen and Ai, Bo and Hu, Rose Qingyang},
  journal={IEEE Transactions on Wireless Communications},
  year={2024},
  publisher={IEEE}
}

@article{wen2023task,
  title={Task-oriented sensing, computation, and communication integration for multi-device edge {AI}},
  author={Wen, Dingzhu and Liu, Peixi and Zhu, Guangxu and Shi, Yuanming and Xu, Jie and Eldar, Yonina C and Cui, Shuguang},
  journal={IEEE Transactions on Wireless Communications},
  volume={23},
  number={3},
  pages={2486--2502},
  year={2023},
  publisher={IEEE}
}

@article{yang2023environment,
  title={Environment semantics aided wireless communications: A case study of {mmWave} beam prediction and blockage prediction},
  author={Yang, Yuwen and Gao, Feifei and Tao, Xiaoming and Liu, Guangyi and Pan, Chengkang},
  journal={IEEE journal on selected areas in communications},
  volume={41},
  number={7},
  pages={2025--2040},
  year={2023},
  publisher={IEEE}
}

@article{wang2025deep,
  title={Deep Learning Assisted {mmWave} Beam Prediction with Flexible Network Architecture},
  author={Wang, Pengyu and Ma, Ke and Bai, Yingshuang and Sun, Chen and Wang, Zhaocheng},
  journal={IEEE Transactions on Wireless Communications},
  year={2025},
  publisher={IEEE}
}

@article{bonfante2021performance,
  title={Performance of predictive indoor {mmWave} networks with dynamic blockers},
  author={Bonfante, Andrea and Giordano, Lorenzo Galati and Macaluso, Irene and Marchetti, Nicola},
  journal={IEEE Transactions on Cognitive Communications and Networking},
  volume={8},
  number={2},
  pages={812--827},
  year={2021},
  publisher={IEEE}
}

@article{wu2022blockage,
  title={Blockage prediction using wireless signatures: Deep learning enables real-world demonstration},
  author={Wu, Shunyao and Alrabeiah, Muhammad and Chakrabarti, Chaitali and Alkhateeb, Ahmed},
  journal={IEEE Open Journal of the Communications Society},
  volume={3},
  pages={776--796},
  year={2022},
  publisher={IEEE}
}

@article{charan2021vision,
  title={Vision-aided {6G} wireless communications: Blockage prediction and proactive handoff},
  author={Charan, Gouranga and Alrabeiah, Muhammad and Alkhateeb, Ahmed},
  journal={IEEE Transactions on Vehicular Technology},
  volume={70},
  number={10},
  pages={10193--10208},
  year={2021},
  publisher={IEEE}
}

@inproceedings{demirhan2022radar,
  title={Radar aided proactive blockage prediction in real-world millimeter wave systems},
  author={Demirhan, Umut and Alkhateeb, Ahmed},
  booktitle={ICC 2022-IEEE International Conference on Communications},
  pages={4547--4552},
  year={2022},
  organization={IEEE}
}

@inproceedings{mehregan2025gcn,
  title={{GCN}-Based Throughput-Oriented Handover Management in Dense {5G} Vehicular Networks},
  author={Mehregan, Nazanin and Robson, E},
  booktitle={2025 21st International Conference on Distributed Computing in Smart Systems and the Internet of Things (DCOSS-IoT)},
  pages={895--902},
  year={2025},
  organization={IEEE}
}

@article{jamshidiha2025power,
  title={Power allocation using spatio-temporal graph neural networks and reinforcement learning},
  author={Jamshidiha, Saeed and Pourahmadi, Vahid and Mohammadi, Abbas and Bennis, Mehdi},
  journal={Wireless Networks},
  volume={31},
  number={2},
  pages={1163--1176},
  year={2025},
  publisher={Springer}
}

@article{corradini2025systematic,
  title={A systematic literature review of spatio-temporal graph neural network models for time series forecasting and classification},
  author={Corradini, Flavio and Gerosa, Flavio and Gori, Marco and Lucheroni, Carlo and Piangerelli, Marco and Zannotti, Martina},
  journal={Neural Networks},
  pages={108269},
  year={2025},
  publisher={Elsevier}
}

@article{jiang2025spatio,
  title={Spatio-temporal {GNN}-Based Cell-Free Massive {MIMO} Network with Maximal Benefit-Cost Ratio},
  author={Jiang, Jing and Li, Yanni and Ye, Yinghui and Feng, Dan and Zhang, Jiayi and Sutthiphan, Worakrin and Niyato, Dusit},
  journal={IEEE Transactions on Vehicular Technology},
  year={2025},
  publisher={IEEE}
}

@article{csahin2023survey,
  title={A survey on over-the-air computation},
  author={{\c{S}}ahin, Alphan and Yang, Rui},
  journal={IEEE Communications Surveys \& Tutorials},
  volume={25},
  number={3},
  pages={1877--1908},
  year={2023},
  publisher={IEEE}
}

@article{yang2023one,
  title={One-bit Byzantine-tolerant distributed learning via over-the-air computation},
  author={Yang, Yuhan and Wu, Youlong and Jiang, Yuning and Shi, Yuanming},
  journal={IEEE Transactions on Wireless Communications},
  volume={23},
  number={6},
  pages={5441--5455},
  year={2023},
  publisher={IEEE}
}

@article{wang2023graph,
  title={A graph neural network learning approach to optimize {RIS}-assisted federated learning},
  author={Wang, Zixin and Zhou, Yong and Zou, Yinan and An, Qiaochu and Shi, Yuanming and Bennis, Mehdi},
  journal={IEEE Transactions on Wireless Communications},
  volume={22},
  number={9},
  pages={6092--6106},
  year={2023},
  publisher={IEEE}
}

@article{yang2023implementing,
  title={Implementing graph neural networks over wireless networks via over-the-air computing: A joint communication and computation framework},
  author={Yang, Yuzhi and Zhang, Zhaoyang and Tian, Yuqing and Jin, Richeng and Huang, Chongwen},
  journal={IEEE Wireless Communications},
  volume={30},
  number={3},
  pages={62--69},
  year={2023},
  publisher={IEEE}
}

@article{lee2023privacy,
  title={Privacy-preserving decentralized inference with graph neural networks in wireless networks},
  author={Lee, Mengyuan and Yu, Guanding and Dai, Huaiyu},
  journal={IEEE Transactions on Wireless Communications},
  volume={23},
  number={1},
  pages={543--558},
  year={2023},
  publisher={IEEE}
}

@article{getu2023making,
  title={Making sense of meaning: A survey on metrics for semantic and goal-oriented communication},
  author={Getu, Tilahun M and Kaddoum, Georges and Bennis, Mehdi},
  journal={IEEE Access},
  volume={11},
  pages={45456--45492},
  year={2023},
  publisher={IEEE}
}

@article{feng2024goal,
  title={Goal-oriented wireless communication resource allocation for cyber-physical systems},
  author={Feng, Cheng and Zheng, Kedi and Wang, Yi and Huang, Kaibin and Chen, Qixin},
  journal={IEEE Transactions on Wireless Communications},
  year={2024},
  publisher={IEEE}
}

@article{zhang2023drl,
  title={{DRL}-driven dynamic resource allocation for task-oriented semantic communication},
  author={Zhang, Haijun and Wang, Hongyu and Li, Yabo and Long, Keping and Nallanathan, Arumugam},
  journal={IEEE Transactions on Communications},
  volume={71},
  number={7},
  pages={3992--4004},
  year={2023},
  publisher={IEEE}
}

@article{wu2024goal,
  title={Goal-oriented semantic communications for robotic waypoint transmission: The value and age of information approach},
  author={Wu, Wenchao and Yang, Yuanqing and Deng, Yansha and Aghvami, A Hamid},
  journal={IEEE Transactions on Wireless Communications},
  year={2024},
  publisher={IEEE}
}

@article{cheng2023resource,
  title={Resource allocation and common message selection for task-oriented semantic information transmission with {RSMA}},
  author={Cheng, Yanyu and Niyato, Dusit and Du, Hongyang and Kang, Jiawen and Xiong, Zehui and Miao, Chunyan and Kim, Dong In},
  journal={IEEE Transactions on Wireless Communications},
  volume={23},
  number={6},
  pages={5557--5570},
  year={2023},
  publisher={IEEE}
}

@INPROCEEDINGS{sionna_ref,
  author={Hoydis, Jakob and Aoudia, Faycal Ait and Cammerer, Sebastian and Nimier-David, Merlin and Binder, Nikolaus and Marcus, Guillermo and Keller, Alexander},
  booktitle={2023 IEEE Globecom Workshops (GC Wkshps)}, 
  title={Sionna {RT}: Differentiable Ray Tracing for Radio Propagation Modeling}, 
  year={2023},
  volume={},
  number={},
  pages={317-321},
  doi={10.1109/GCWkshps58843.2023.10465179}}

@inproceedings{wu2022lidar,
  title={{LiDAR}-aided mobile blockage prediction in real-world millimeter wave systems},
  author={Wu, Shunyao and Chakrabarti, Chaitali and Alkhateeb, Ahmed},
  booktitle={2022 IEEE Wireless Communications and Networking Conference (WCNC)},
  pages={2631--2636},
  year={2022},
  organization={IEEE}
}

@ARTICLE{10843389,

  author={Longhi, Nicolò and Amorosa, Lorenzo Mario and Cavallero, Sara and Buracchini, Enrico and Verdone, Roberto},

  journal={IEEE Open Journal of the Communications Society}, 

  title={{5G} Architectures Enabling Remaining Useful Life Estimation for Industrial {IoT}: A Holistic Study}, 

  year={2025},

  volume={6},

  number={},

  pages={1016-1029},

  doi={10.1109/OJCOMS.2025.3530094}}

@inproceedings{pezone2022goal,
  title={Goal-oriented communication for edge learning based on the information bottleneck},
  author={Pezone, Francesco and Barbarossa, Sergio and Di Lorenzo, Paolo},
  booktitle={ICASSP 2022-2022 IEEE International Conference on Acoustics, Speech and Signal Processing (ICASSP)},
  pages={8832--8836},
  year={2022},
  organization={IEEE}
}

@INPROCEEDINGS{11073510,
  author={Amorosa, Lorenzo Mario and Spampinato, Leonardo and Buratti, Chiara and Verdone, Roberto},
  booktitle={IEEE EUROCON 2025 - 21st International Conference on Smart Technologies}, 
  title={Goal-Oriented Uplink Scheduling Requests in Wireless Networks via Graph Neural Networks}, 
  year={2025},
  volume={},
  number={},
  pages={1-6},
  doi={10.1109/EUROCON64445.2025.11073510}}

@INPROCEEDINGS{ctde1,

  author={Amorosa, Lorenzo Mario and Gao, Zhan and Chahoud, Tony and Verdone, Roberto and Gündüz, Deniz},

  booktitle={2025 IEEE International Conference on Machine Learning for Communication and Networking (ICMLCN)}, 

  title={Decentralized {GNN}-based Power Allocation with Varying Network Density}, 

  year={2025},

  volume={},

  number={},

  pages={1-6},

  doi={10.1109/ICMLCN64995.2025.11139371}}

@INPROCEEDINGS{ctde2,
  author={Amorosa, Lorenzo Mario and Skocaj, Marco and Verdone, Roberto and Gündüz, Deniz},
  booktitle={ICC 2024 - IEEE International Conference on Communications}, 
  title={Multi-Agent Reinforcement Learning for Power Control in Wireless Networks via Adaptive Graphs}, 
  year={2024},
  volume={},
  number={},
  pages={2968-2973},
  doi={10.1109/ICC51166.2024.10622170}}

@article{mahmood2019time,
  title={Time synchronization in {5G} wireless edge: Requirements and solutions for critical-{MTC}},
  author={Mahmood, Aamir and Ashraf, Muhammad Ikram and Gidlund, Mikael and Torsner, Johan and Sachs, Joachim},
  journal={IEEE Communications Magazine},
  volume={57},
  number={12},
  pages={45--51},
  year={2019},
  publisher={IEEE}
}

@inproceedings{abari2015airshare,
  title={{AirShare}: Distributed coherent transmission made seamless},
  author={Abari, Omid and Rahul, Hariharan and Katabi, Dina and Pant, Mondira},
  booktitle={2015 IEEE Conference on Computer Communications (INFOCOM)},
  pages={1742--1750},
  year={2015},
  organization={IEEE}
}

@inproceedings{csahin2022over,
  title={Over-the-air computation over balanced numerals},
  author={{\c{S}}ahin, Alphan and Yang, Rui},
  booktitle={2022 IEEE Globecom Workshops (GC Wkshps)},
  pages={347--352},
  year={2022},
  organization={IEEE}
}

@article{wang2022amplify,
  title={Amplify-and-forward relaying for hierarchical over-the-air computation},
  author={Wang, Feng and Xu, Jie and Lau, Vincent KN and Cui, Shuguang},
  journal={IEEE Transactions on Wireless Communications},
  volume={21},
  number={12},
  pages={10529--10543},
  year={2022},
  publisher={IEEE}
}

@article{chen2018over,
  title={Over-the-air computation for {IoT} networks: Computing multiple functions with antenna arrays},
  author={Chen, Li and Zhao, Nan and Chen, Yunfei and Yu, F Richard and Wei, Guo},
  journal={IEEE Internet of Things journal},
  volume={5},
  number={6},
  pages={5296--5306},
  year={2018},
  publisher={IEEE}
}

@ARTICLE{letaief2019roadmap,
  author={Letaief, Khaled B. and Chen, Wei and Shi, Yuanming and Zhang, Jun and Zhang, Ying-Jun Angela},
  journal={IEEE Communications Magazine}, 
  title={The Roadmap to {6G}: {AI} Empowered Wireless Networks}, 
  year={2019},
  volume={57},
  number={8},
  pages={84-90},
  doi={10.1109/MCOM.2019.1900271}}

@INPROCEEDINGS{9567793,
  author={Strinati, Emilio Calvanese and Belot, Didier and Falempin, Alexis and Doré, Jean-Baptiste},
  booktitle={ESSCIRC 2021 - IEEE 47th European Solid State Circuits Conference (ESSCIRC)}, 
  title={Toward {6G}: From New Hardware Design to Wireless Semantic and Goal-Oriented Communication Paradigms}, 
  year={2021},
  volume={},
  number={},
  pages={275-282},
  doi={10.1109/ESSCIRC53450.2021.9567793}}

@incollection{frey2022over,
  title={Over-the-air computation for distributed machine learning and consensus in large wireless networks},
  author={Frey, Matthias and Bjelakovi{\'c}, Igor and Sta{\'n}czak, S{\l}awomir},
  booktitle={Compressed Sensing in Information Processing},
  pages={401--434},
  year={2022},
  publisher={Springer}
}

@article{ren2023survey,
  title={A survey on collaborative {DNN} inference for edge intelligence},
  author={Ren, Wei-Qing and Qu, Yu-Ben and Dong, Chao and Jing, Yu-Qian and Sun, Hao and Wu, Qi-Hui and Guo, Song},
  journal={Machine Intelligence Research},
  volume={20},
  number={3},
  pages={370--395},
  year={2023},
  publisher={Springer}
}

@ARTICLE{9170818,
  author={Wu, Wen and Yang, Peng and Zhang, Weiting and Zhou, Conghao and Shen, Xuemin},
  journal={IEEE Transactions on Industrial Informatics}, 
  title={Accuracy-Guaranteed Collaborative {DNN} Inference in Industrial {IoT} via Deep Reinforcement Learning}, 
  year={2021},
  volume={17},
  number={7},
  pages={4988-4998},
  doi={10.1109/TII.2020.3017573}}

@INPROCEEDINGS{10226176,
  author={Sagduyu, Yalin E. and Ulukus, Sennur and Yener, Aylin},
  booktitle={IEEE INFOCOM 2023 - IEEE Conference on Computer Communications Workshops (INFOCOM WKSHPS)}, 
  title={Age of Information in Deep Learning-Driven Task-Oriented Communications}, 
  year={2023},
  volume={},
  number={},
  pages={1-6},
  doi={10.1109/INFOCOMWKSHPS57453.2023.10226176}}

@ARTICLE{10538293,
  author={Wang, Zhibin and Zhao, Yapeng and Zhou, Yong and Shi, Yuanming and Jiang, Chunxiao and Letaief, Khaled B.},
  journal={IEEE Internet of Things Journal}, 
  title={Over-the-Air Computation for {6G}: Foundations, Technologies, and Applications}, 
  year={2024},
  volume={11},
  number={14},
  pages={24634-24658},
  doi={10.1109/JIOT.2024.3405448}}

@ARTICLE{10707077,
  author={Lu, Yang and Li, Yuhang and Zhang, Ruichen and Chen, Wei and Ai, Bo and Niyato, Dusit},
  journal={IEEE Wireless Communications}, 
  title={Graph Neural Networks for Wireless Networks: Graph Representation, Architecture and Evaluation}, 
  year={2025},
  volume={32},
  number={1},
  pages={150-156},
  doi={10.1109/MWC.006.2400131}}

@incollection{pardalos1998graph,
  title={The graph coloring problem: A bibliographic survey},
  author={Pardalos, Panos M and Mavridou, Thelma and Xue, Jue},
  booktitle={Handbook of Combinatorial Optimization: Volume1--3},
  pages={1077--1141},
  year={1998},
  publisher={Springer}
}

@ARTICLE{8870236,
  author={Zhu, Guangxu and Wang, Yong and Huang, Kaibin},
  journal={IEEE Transactions on Wireless Communications}, 
  title={Broadband Analog Aggregation for Low-Latency Federated Edge Learning}, 
  year={2020},
  volume={19},
  number={1},
  pages={491-506},
  doi={10.1109/TWC.2019.2946245}}
\endgroup

\end{document}